\documentclass[aps,prb,twocolumn,superscriptaddress,amsmath,amssymb,amsfonts,floatfix]{revtex4-2}

\usepackage[utf8]{inputenc}
\usepackage[english]{babel}
\usepackage[T1]{fontenc}
\usepackage{float}
\usepackage{graphicx}
\usepackage{dcolumn}
\usepackage{amsmath}
\usepackage{amssymb}
\usepackage{bm}
\usepackage[urlcolor=blue,colorlinks=true,citecolor=red,linkcolor=red,pdfstartview={FitH},bookmarks=false]{hyperref}
\usepackage{xcolor}
\usepackage{listings}
\usepackage{enumitem}
\usepackage{epstopdf}
\usepackage[normalem]{ulem}

\def\vnabla{{\boldsymbol{\nabla}}}

\def\vA{{\bf A}}
\def\vAext{{\bm{A}_{\mathrm{ext}}}}
\def\vAind{{\bm{A}_{\mathrm{ind}}}}

\def\vBext{{\bm{B}_{\mathrm{ext}}}}
\def\vBind{{\bm{B}_{\mathrm{ind}}}}

\def\Bind{{B_{\mathrm{ind}}}}

\def\BcTwo{{B_{\mathrm{c},2}}}

\def\vzhat{{\hat{\bm{z}}}}

\def\Phiext{{\Phi_{\mathrm{ext}}}}

\def\vj{{\bf j}}

\def\vvF{{\bf v}_{\mathrm{F}}}
\def\vF{v_{\mathrm{F}}}

\def\vps{{\bf p}_{\mathrm{s}}}

\def\vpF{{\bf p}_{\mathrm{F}}}
\def\vpFp{{\bf p}^{\prime}_{\mathrm{F}}}
\def\vpFpp{{\bf p}^{\prime\prime}_{\mathrm{F}}}
\def\vR{{\bf R}}

\def\RCV{\mathcal{R}_{\mathrm{CV}}}
\def\pF{p_{\mathrm{F}}}

\def\EF{E_{\mathrm{F}}}
\def\lambdaF{\lambda_{\mathrm{F}}}
\def\NF{N_{\mathrm{F}}}
\def\FS{{\mathrm{FS}}}
\def\thetaF{{\theta_{\mathrm{F}}}}

\def\kB{k_{\mathrm{B}}}

\def\Tc{T_{\mathrm{c}}}
\def\Tcdxy{T_{\mathrm{c}}^{d_{xy}}}
\def\Tcdxtyt{T_{\mathrm{c}}^{d_{x^2-y^2}}}

\def\Omegac{\Omega_{\mathrm{c}}}

\def\Dxtyt{{\Delta_{d_{x^2-y^2}}}}
\def\Dxy{{\Delta_{d_{xy}}}}

\def\etaxtyt{{\eta_{d_{x^2-y^2}}}}
\def\etaxy{{\eta_{d_{xy}}}}
\def\etapm{{\eta_{\pm}}}
\def\chixtyt{{\chi_{d_{x^2-y^2}}}}
\def\chixy{{\chi_{d_{xy}}}}
\def\chipm{{\chi_{\pm}}}

\def\Dplus{{\Delta_{+}}}
\def\Dminus{{\Delta_{-}}}
\def\Dpm{{\Delta_{\pm}}}

\def\Lz{{\hat{L}_z}}
\def\Lzorb{{\hat{L}_z^{\mathrm{orb}}}}
\def\Lzcm{{\hat{L}_z^{\mathrm{c.m.}}}}

\def\xieff{{\xi_{\mathrm{eff}}}}
\def\alphaxy{{\alpha_{xy}}}

\allowdisplaybreaks

\begin{document}

\title{Robust and tunable coreless vortices and fractional vortices in chiral $d$-wave superconductors}

\author{P. Holmvall}
\affiliation{Department of Physics and Astronomy,
	Uppsala University, Box 516, S-751 20, Uppsala, Sweden}
\author{N. Wall-Wennerdal}
\affiliation{Department of Microtechnology and Nanoscience - MC2, Chalmers University of Technology, SE-412 96 G\"oteborg, Sweden}
\author{A. M. Black-Schaffer}
\affiliation{Department of Physics and Astronomy,
	Uppsala University, Box 516, S-751 20, Uppsala, Sweden}

\date{\today}

\begin{abstract}
Chiral $d$-wave superconductivity has recently been proposed in a wide range of materials based on both experiment and theoretical works. Chiral superconductors host a finite Chern number set by the winding of the superconducting order parameter and associated topologically protected chiral edge modes. However, the chiral edge currents and orbital angular momentum (OAM) generated by the edge modes are not topologically protected and another, more robust, experimental probe is therefore needed to facilitate experimental verification of chiral $d$-wave superconductors. We have recently shown the appearance of quadruply quantized coreless vortices (CVs) in chiral $d$-wave superconductors, consisting of a closed domain wall decorated with eight fractional vortices, and generating a smoking-gun signature of the Chern number, chirality, and the superconducting pairing symmetry [P. Holmvall and A. M. Black-Schaffer, arXiv:2212.08156 (2023)]. Specifically, the CV spontaneously breaks axial symmetry for parallel chirality and vorticity, with a signature appearing directly in the local density of states (LDOS) measurable with scanning-tunneling spectroscopy (STS). In this work, we first demonstrate a strong tunability of the CV size and shape directly reflected in the LDOS and then show that the LDOS signature is robust in the presence of regular Abrikosov vortices, strong confinement, system and normal-state anisotropy, different Fermi surfaces (FSs), non-degenerate order parameters, and even non-magnetic impurities. In conclusion, our work establishes CVs as a tunable and robust signature of chiral $d$-wave superconductivity.

\end{abstract}

\maketitle

\section{Introduction}
\label{sec:intro}

Two of the most outstanding issues in condensed matter physics are the direct identification of the superconducting pairing symmetry in unconventional superconductors and of the topological invariant in topologically non-trivial materials. These difficulties severely limit the ability to correctly interpret experiments and the applicability of newly discovered superconducting and topological materials. In few other systems is this as problematic as in multi-component superconductors, especially chiral superconductors, where both topology and superconducting symmetry need to be identified. Theoretically, chiral superconductors, and more generally chiral superfluids, are characterized by a non-trivial topology \cite{Volovik:2003,Kallin:2016,Mizushima:2016,Volovik:2019,Volovik:2020} and a discretely degenerate ground state that spontaneously breaks time-reversal symmetry \cite{Sigrist:1991}. 
They belonging to the class of integer quantum Hall systems \cite{Volovik:1988,Volovik:1992,Read:2000} with a finite Chern number generated by the winding of the superfluid order parameter \cite{Volovik:1997,Schnyder:2008,Hasan:2010,Qi:2011,Sauls:2011,Tanaka:2012,Graf:2013,Black-Schaffer:2014:b}, and with topologically protected chiral edge modes generating spontaneous surface currents and orbital angular momentum (OAM) \cite{Anderson:1961,Leggett:1975,McClure:1979,Sauls:1994,Kita:1998,Stone:2004}.

The topology and symmetry breaking of a chiral superfluid are predicted to generate a range of interesting properties \cite{Volovik:2003,Kallin:2016,Mizushima:2016,Volovik:2019,Volovik:2020,Heikkinen:2021,Volkov:2023:PRL,Volkov:2023:PRB,Cadorim:2022:arxiv}, such as the existence of domain walls \cite{Volovik:2003,Beri:2010,Becerra:2016:a,Awoga:2017}, states with non-Abelian statistics \cite{Volovik:1999,Ivanov:2001,Mercado:2022,Margalit:2022,Huang:2023,Li:2023}, proposed as a platform for topological quantum computing \cite{Beenakker:2013,Sato:2017}, and fascinating vortex defects without analogues in single-component superfluids \cite{Volovik:2003,Kallin:2016,Mizushima:2016,Volovik:2019,Volovik:2020,Volovik:2016:a}. A prime example is the continuous ``coreless vortex'' (CV), which due to its multi-component structure is non-singular with finite superfluid order parameter everywhere. CVs have primarily been studied in superfluid ${^{3}\mathrm{He}}$ \cite{Mermin:1976,Chechetkin:1976,Anderson:1977, Ho:1978,Hakonen:1982,Seppala:1983,Seppala:1984,Thuneberg:1986,Salomaa:1987,Parts:1995:a,Ruutu:1997,Lounasmaa:1999,Blaauwgeers:2000,Walmsley:2003,Takagi:2005,Rantanen:2023}. In superconductors, CVs have so far mainly been discussed in the context of spin-triplet chiral $p$-wave \cite{Sauls:2009,Garaud:2012,Takamatsu:2013,Garaud:2015,Garaud:2016,Zhang:2016,Becerra:2016:b,Zyuzin:2017,Zha:2020,Chai:2021,Krohg:2021,Tokiwa:2023}, with analogous states discussed for various multi-band superconductors and other multi-component condensates \cite{Babaev_Faddeev:2002,Garaud:2011,Garaud:2013,Winyard:2019,Zhang:2020,Benfenati:2023}. In the superconducting scenario, the CV essentially consists of a closed domain wall, along which the vorticity enters as fractional vortices, such that the total superconducting order parameter is non-singular and finite everywhere. Fractional vortices have been studied extensively over the years \cite{Volovik:1976,Sigrist:1989,Sigrist:1991,Volovik:2000,Tsuei:2000,Tsuei:2000:b,Babaev:2002,Babaev:2004,Babaev:2007,Chibotaru:2007,Chibotaru:2010,Silaev:2011,Pina:2012,Mineev:2013,Tanaka:2023}, and were recently experimentally observed in superfluids \cite{Autti:2016} and superconductors \cite{Iguchi:2023}. Similar to a regular Abrikosov vortex \cite{Abrikosov:1957}, the CV is stabilized by its reduction of the kinetic energy in an external magnetic field. But importantly, unlike an Abrikosov vortex (or a giant vortex \cite{Fink:1966,Moshchalkov:1996,Schweigert:1998,Kanda:2004,Golubovic:2005,Cren:2011}), the CV by definition has no normal core, and therefore avoids the usual energy penalty associated with lost condensation in the core.

Recent decades have seen an intense search for experimental realizations of chiral superconductors due to their many interesting properties and proposed applications \cite{Kallin:2012,Kallin:2016,Mizushima:2016}. The hunt for chiral superconductivity has mainly focused on spin-triplet chiral $p$-wave and $f$-wave superconductivity \cite{Mackenzie:2003,Kallin:2012,Kallin:2016,Suh:2020,Duan:2021,Bae:2021,Hayes:2021}, and their similarities with superfluid $^{3}\mathrm{He}\text{\ensuremath{-}}\mathrm{A}$ \cite{Volovik:2003,Kallin:2016,Mizushima:2016,Volovik:2019,Volovik:2020}. Interestingly, multiple proposals of spin-singlet chiral $d$-wave superconductivity have more recently emerged based both on theory and experiments in a range of materials, such as twisted bilayer cuprates \cite{Can:2021:a,Can:2021:b}, twisted bilayer graphene \cite{Venderbos:2018,Su:2018,Fidrysiak:2018,Xu:2018,Kennes:2018,Liu:2018,Gui:2018,Wu:2019,Fischer:2021}, ${\mathrm{Sn/Si}}(111)$ \cite{Ming:2023}, $\textrm{SrPtAs}$ \cite{Biswas:2013,Fischer:2014,Ueki:2019,Ueki:2020}, $\textrm{LaPt$_3$P}$ \cite{Biswas:2021}, ${\mathrm{Bi/Ni}}$ \cite{Gong:2017,Hosseinabadi:2019} and $\mathrm{U}{\mathrm{Ru}}_{2}{\mathrm{Si}}_{2}$ \cite{Kasahara:2007,Kasahara:2009,Shibauchi:2014,Iguchi:2021}. Chiral $d$-wave superconductivity was recently also proposed as a route to topologically protected quantum computing \cite{Mercado:2022,Margalit:2022,Huang:2023,Li:2023}. The exact identification of the superconducting pairing symmetry is however still highly debated in these proposed chiral superconductors. This is further hampered by the fact that typical fingerprints of chiral superconductivity, namely the chiral edge currents and OAM are not topologically protected \cite{Volovik:1988,Black-Schaffer:2012,Nie:2020}, and may even often vanish for pairing symmetries except for $p$-wave \cite{Huang:2014,Tada:2015,Volovik:2015:b,Suzuki:2016,Ojanen:2016,Wang:2018,Tada:2018,Nie:2020,Sugiyama:2020}. In addition, it is quite unknown how the higher Chern number and angular momentum of chiral $d$-wave superconductors influence the vortex physics and CVs.

In an earlier work we have demonstrated that CVs naturally emerge as a `quadruple-quantum vortex' in spin-singlet chiral $d$-wave superconductors and that they, most importantly, act as a smoking-gun signature of chirality, pairing symmetry, and Chern number \cite{Holmvall:2023:arxiv}. These signatures were demonstrated directly in the local density of states (LDOS) and indirectly in the area-averaged orbital magnetic moment, the former measurable with e.g.~scanning tunneling spectroscopy (STS) and scanning tunneling microscopy (STM) \cite{Hess:1989,Renner:1991,Maggio-Aprile:1995,Yazdani:1997,Pan:2000a,Pan:2000b,Hoffman:2002,Hoffman:2002:b,Guillamon:2008,Roditchev:2015,Berthod:2017}, and the latter with various magnetometry setups \cite{Geim:1997,Bolle:1999,Tsuei:2000,Morelle:2004,Kirtley:2005,Khotkevych:2008,Bleszynski-Jayich:2009,Kokubo:2010,Bert:2011,Jang:2011,Vasyukov:2013,Curran:2014,Ge:2017,Persky:2022}. The signatures were shown to be fundamentally related to the existence of inequivalent CVs in opposite magnetic field directions (or equivalently opposite chiralities), due to either a parallel or antiparallel vorticity and chirality, and which are also completely different from regular Abrikosov vortices.

In this complementary work, we demonstrate a strong tunability of the CV size and shape, also directly reflected in e.g.~the LDOS. Furthermore, we provide extensive data that demonstrate a strong robustness of the results for a range of realistic models, over extensive parameter ranges, and in the presence of additional vortices or disorder. Overall, we relate the robustness of the experimental signatures of chirality, pairing symmetry, and Chern number, to the fact that they fundamentally stem from the parallel versus antiparallel alignment of vorticity and chirality, which are both topologically protected. In contrast, a non-chiral superconductor lacks this alignment possibility, since it lacks chirality. Our work therefore establishes CVs as a robust signature of spin-singlet chiral $d$-wave superconductivity, and furthermore the realization of fractional vortices in these materials.

This work is organized as follows. In Sec.~\ref{sec:methods} we summarize our model and methods, and describe basic properties of chiral $d$-wave superconductors.
In Sec.~\ref{sec:coreless_vortices} we introduce the basic properties of CVs, also discussing their overall stability and formation.
In Sec.~\ref{sec:cv_size} we demonstrate the large tunability of the CV size due to thermodynamics and electrodynamic interactions.
Similarly, we study the interaction between CVs and other vortices in Sec.~\ref{sec:cv_v_interaction} and the behavior of CVs in confinement in Sec.~\ref{sec:non_circular_grains}, again demonstrating a tunability of both the CV size and shape as well as establishing strong robustness of CVs. 
We further demonstrate robustness against more general and anisotropic Fermi surfaces (FSs) in Sec.~\ref{sec:non_circular_fs}, non-degenerate order parameter components in Sec.~\ref{sec:non_degeneracy}, and non-magnetic impurities in Sec.~\ref{sec:impurities}. Finally in Sec.~\ref{sec:conclusions} we briefly summarize our results.

\section{Model and methods}
\label{sec:methods}
In this section, we describe our model for a spin-singlet chiral $d$-wave superconductor and summarize our methods. In particular, we use the quasiclassical theory of superconductivity \cite{Eilenberger:1968,Larkin:1969,Serene:1983,Shelankov:1985,Nagato:1993,Eschrig:1994,Schopohl:1995,Schopohl:1998,Eschrig:1999,Eschrig:2000,Eschrig:2009,Grein:2013,Seja:2022:FEM}, and perform self-consistent numerical calculations using the open-source framework SuperConga \cite{SuperConga:2023}.

\subsection{Model}
\label{sec:methods:model}
We consider weak-coupling superconductivity in equilibrium and in two dimensions (2D), assuming spin-degeneracy, all appropriate for a quantitative description of a spin-singlet $d$-wave superconductor. We start by studying clean superconductors shaped like discs, with an electron-doped and circular Fermi surface (FS). We then relax all these assumptions by studying systems with either different discrete rotational symmetries or completely irregular shapes, as well as hole-doped and anisotropic FSs. We also consider non-degenerate order parameters, as well as dirty superconductors with non-magnetic impurities. For the specific setup, we align the superconducting plane with the $xy$-axes and use a perpendicular (orbital) external magnetic flux density $\vBext = (\Phiext/\mathcal{A})\vzhat$ with homogeneous flux $\Phiext$ across the system area $\mathcal{A}$ to induce vortex states. We assume type-II superconductivity appropriate for most non-elemental or unconventional superconductors, but consider different penetration depths $\lambda_0 \in [2,\infty)$, via the Ginzburg-Landau coefficient $\kappa \equiv \lambda_0/\xi_0$. The penetration depth sets the length scale and strength of flux screening, defined by $\lambda_0^{-2} = 4\pi e^2 \vF^2\NF/c^2$, with elementary charge $e = -|e|$, Fermi velocity $\vF$ on the FS, normal-state density of state $\NF$ on the FS (per spin), and speed of light $c$. Here, our natural length unit is $\xi_0 \equiv \hbar\vF/(2\pi\kB\Tc)$, sometimes referred to as an effective superconducting coherence length over which superconductivity spatially varies, with Planck constant $\hbar$, Boltzmann constant $\kB$, and superconducting transition temperature $\Tc$. We study superconducting systems with a diameter or side length $\mathcal{D} \in [20,300]\xi_0$, for different temperatures $T \in [0.01,0.99]\Tc$, and external fluxes $\Phiext\in[-15,15]\Phi_0$ with flux quantum $\Phi_0 \equiv hc/2|e|$. We keep all parameters fixed during the self-consistency simulations.

We perform our numerical simulations using the open-source framework SuperConga \cite{SuperConga:2023}, which is a state-of-the-art implementation of the quasiclassical theory of superconductivity \cite{Eilenberger:1968,Larkin:1969,Serene:1983,Shelankov:1985,Nagato:1993,Eschrig:1994,Schopohl:1995,Schopohl:1998,Eschrig:1999,Eschrig:2000,Eschrig:2009,Grein:2013,Seja:2022:FEM}, running on graphics processing units (GPUs), and with extensive documentation and unit testing \cite{SuperConga:documentation,SuperConga:repository}. SuperConga solves self-consistently \footnote{Our criteria for self-consistency is that the global relative error should be ${<10^{-7}}$ for $\Delta$, $\vA$, the free energy $\Omega$, and charge-current density $\vj$.} for both the superconducting order parameter $\Delta(\vpF,\vR)$ and vector potential $\vA(\vR)$ via the gap equation and Maxwell's equations, respectively. Here,  $\vpF=\pF(\cos\thetaF,\sin\thetaF)$ is the Fermi momentum with angle $\thetaF$ on the FS, while $\vR = R(\cos\phi,\sin\phi)$ is the in-plane center-of-mass coordinate with polar angle $\phi$. SuperConga also solves for impurity self-energies self-consistently using the well-established $t$-matrix approach \cite{Graf:1996}.

\subsection{Quasiclassical theory of superconductivity}
\label{sec:methods:quasiclassics}
Many materials exhibit a clear separation between the superconducting gap $|\Delta|$ and other relevant energy scales, such as the Fermi energy $\EF$. Consequently, the superconducting coherence length $\xi_0$ typically becomes much larger than the atomic length scale $a_0$ and Fermi wavelength $\lambda_F$. In such materials, the low-energy (long-wavelength) physics can often to a very good approximation be separated from the high-energy (short-wavelength) physics. The quasiclassical theory of superconductivity exploits this via a controlled expansion in the resulting small parameters, e.g.~$|\Delta|/\EF$, $T/\Tc$, and $\lambdaF/\xi_0$, with leading-order terms describing the low-energy bands close to the FS \cite{Eilenberger:1968,Larkin:1969,Serene:1983,Shelankov:1985,Nagato:1993,Eschrig:1994,Schopohl:1995,Schopohl:1998,Eschrig:1999,Eschrig:2000,Eschrig:2009,Grein:2013,Seja:2022:FEM}. Higher-energy corrections can still be inserted from full microscopic theory, e.g.~by using microscopic boundary conditions \cite{Zaitsev:1984,Kieselmann:1987,Millis:1988,Eschrig:2000,Shelankov:2000,Fogelstrom:2000,Zhao:2004,Eschrig:2009}.

The low-energy expansion results in quasiclassical propagators, which we express in Nambu (particle-hole) space as
\begin{align}
\label{eq:model:green_function}
\hat{g}(\vpF, \vR; z) = 
\begin{pmatrix}
    g(\vpF, \vR; z) & f(\vpF, \vR; z)\\
    -\tilde{f}(\vpF, \vR; z) & \tilde{g}(\vpF, \vR; z)
\end{pmatrix},
\end{align}
with quasiparticle propagator $g(\vpF, \vR; z)$ and anomalous pair propagator $f(\vpF, \vR; z)$, where ``tilde'' denotes particle-hole conjugation $\tilde{\alpha}(\vpF, \vR; z) = \alpha^*(-\vpF, \vR; -z^*)$.  Here, $z$ is the quasiparticle energy associated with the corresponding propagator, and is generally complex valued. Specifically, the retarded propagators $g^{\mathrm{R}}(\vpF,\vR;\varepsilon)$ are used for spectral quantities, evaluated at $z^{\mathrm{R}} \equiv \varepsilon + i\delta$ with real energy $\varepsilon$ and small positive broadening $\delta$. For all other quantities, we use the Matsubara propagators $g^{\mathrm{M}}(\vpF,\vR;\varepsilon_n)$ and $f^{\mathrm{M}}(\vpF,\vR;\varepsilon_n)$ in terms of the Matsubara energies $z^{\mathrm{M}} \equiv i\varepsilon_n = i\pi\kB T(2n+1)$, with integer $n$ \cite{Matsubara:1955,Bruus:2004,Rammer:2007,Ozaki:2007,Kopnin:2009,Mahan:2013}. The propagators in Eq.~(\ref{eq:model:green_function}) are obtained via the Eilenberger equation \cite{Eilenberger:1968}
\begin{align}
    \nonumber
    0 = & i\hbar\vvF\cdot\boldsymbol{\nabla}\hat{g}(\vpF, \vR; z)\\
    \label{eq:model:eilenberger}
    & + \left[z\hat{\tau}_3 - \hat{h}(\vpF, \vR; z),\hat{g}(\vpF, \vR; z)\right] ,
\end{align}
together with the normalization condition $\hat{g}^2 = -\pi^2\hat{1}$, where $\hat{h}$ is the self energy and $\hat{\tau}_i$ the Pauli matrices in Nambu space. The self energy in Nambu space is
\begin{align}
    \nonumber
    \hat{h}(\vpF, \vR; z) & = \hat{\Sigma}(\vpF, \vR; z) + \hat{\Delta}(\vpF, \vR)\\
    \label{eq:model:self_energy}
    & = \begin{pmatrix}
    \Sigma(\vpF, \vR; z) & \Delta(\vpF, \vR) \\
    \tilde{\Delta}(\vpF, \vR) & \tilde{\Sigma}(\vpF, \vR; z)
    \end{pmatrix},
\end{align}
with mean-field superconducting order parameter $\Delta(\vpF, \vR)$, while the diagonal part in the present work is
\begin{align}
    \label{eq:model:self_energy:diagonal}
    \hat{\Sigma}(\vpF, \vR; z) = \hat{\Sigma}_{\mathrm{flux}}(\vR) + \hat{\Sigma}_{\mathrm{imp}}(\vpF, \vR; z),
\end{align}
capturing electrodynamic interactions via $\hat{\Sigma}_{\mathrm{flux}}(\vR)$ (described further below) and impurity scattering via $\hat{\Sigma}_{\mathrm{imp}}(\vpF, \vR; z)$ (described in Sec.~\ref{sec:impurities}).
We parametrize the even-parity spin-singlet order parameter $\Delta(\vpF,\vR)$ via
\begin{align}
    \label{eq:model:order_parmeter:irreducible_representation}
    \Delta(\vpF,\vR) = \sum_\Gamma|\Delta_\Gamma(\vR)|e^{i\chi_\Gamma(\vR)}\eta_\Gamma(\vpF),
\end{align}
where $\Gamma$ labels the irreducible representations of the crystallographic point group and the basis function $\eta_\Gamma(\vpF)$ encodes the pairing symmetry on the FS \cite{Yip:1993}, also related to the attractive pairing interaction $V$ via
\begin{align}
    \label{eq:model:pairing_interaction}
    V(\vpF,\vpF^\prime)=\sum_{\Gamma} V_\Gamma \eta_\Gamma(\vpF)\eta^\dagger_\Gamma(\vpF^\prime).
\end{align}
Here, $V_\Gamma$ is the pairing strength of the respective symmetry channel. We self-consistently compute $\Delta(\vpF,\vR)$ via the superconducting gap equation
\begin{align}
    \label{eq:model:gap_equation}
    \Delta(\vpF,\vR) = \NF \kB  T\sum_n^{|\varepsilon_n|<\Omegac} \big\langle V(\vpF,\vpF^\prime)\,f(\vpF^\prime,\vR;\varepsilon_n)
                   \big\rangle_{\vpF^\prime},
\end{align}
with cutoff energy $\Omegac$ \cite{Grein:2013}, and FS average \cite{Graf:1993,Wennerdal:2020}
\begin{align}
    \label{eq:model:fs_average}
    \left\langle \dots \right\rangle_{\vpF} = \frac{1}{\NF} \oint_\FS \frac{\mathrm{d} p_\mathrm{F}}{(2\pi\hbar)^2|\vvF(\vpF)|} (\dots).
\end{align}
The electrodynamics is modelled via
\begin{align}
    \label{eq:model:self_energy:flux}
    \hat{\Sigma}_{\mathrm{flux}}(\vpF, \vR) = -\frac{e}{c}\vvF(\vpF)\cdot\vA(\vR)\hat{\tau}_3,
\end{align}
where $\vA(\vR) = \vAext(\vR) + \vAind(\vR)$ is the magnetic vector potential. It is related to the external (ext) magnetic-flux density via Maxwell's equation $\vBext(\vR) = \vnabla\times\vAext(\vR)$, and to the induced (ind) magnetic-flux density $\vBind(\vR)$ (i.e.~screening) from the total charge-current density $\vj(\vR)$ via Amp{\`e}re's law
\begin{align}
    \label{eq:ampere}
    \frac{4\pi}{c}\vj(\vR) = \vnabla\times\vBind(\vR) = \vnabla\times\vnabla\times\vAind(\vR).
\end{align}
We compute $\vj(\vR)$ via
\begin{align}
    \label{eq:current}
    \vj(\vR) = 2e \NF \kB T\sum_{n}^{|\varepsilon_n|<\Omegac} \left\langle \vvF(\vpF)\, g^{\mathrm{M}}(\vpF,\vR;\varepsilon_n) \right\rangle_{\vpF}.
\end{align}
We further compute the LDOS via
\begin{align}
    \label{eq:model:ldos}
    N(\vR; \varepsilon) = -\frac{2\NF}{\pi} \left\langle \operatorname{Im}\left[ g^{\mathrm{R}}(\vpF, \vR; \varepsilon)\right] \right\rangle_{\vpF}.
\end{align}
Finally, we note that the quasiparticle energies are effectively Doppler shifted by the vector potential and any phase gradients \cite{Kubert:1998,Kohen:2006,Hakansson:2015,Holmvall:thesis:2019}, seen by applying a unitary gauge transformation to the Eilenberger equation as in e.g. Refs.~\cite{Sharma:2020,Holmvall:2018b,SuperConga:2023}, modifying Eq.~(\ref{eq:model:self_energy:flux}), $\Sigma_{\mathrm{flux}}(\vpF, \vR) \to \vvF(\vpF)\cdot\vps(\vR)$ with the gauge-invariant superfluid momentum (superflow)
\begin{align}
    \label{eq:superflow}
    \vps(\vR) = \frac{\hbar}{2}\vnabla\chi(\vR) - \frac{e}{c}\vA(\vR).
\end{align}
This allows phase gradients and vector potentials to be treated on an equal footing, and leads to the Doppler shifted quasiparticle energy $z_p = z - \vvF(\vpF)\cdot\vps(\vR)$ in the Eilenberger equation Eq.~(\ref{eq:model:eilenberger}) \cite{Xu:1995,Holmvall:2019,Holmvall:2020}, thus also influencing the LDOS in Eq.~(\ref{eq:model:ldos}).

\subsection{Chiral superconductivity}
\label{sec:methods:chiral}
We consider spin-singlet chiral $d$-wave superconductivity, modelled using an attractive pair potential for the two irreducible $d$-wave representations $\Gamma \in \{d_{x^2-y^2}, d_{xy}\}$ with $\etaxtyt(\thetaF)=\sqrt{2}\cos(2\thetaF)$ and $\etaxy(\thetaF)=\sqrt{2}\sin(2\thetaF)$. Following the notation in Eq.~(\ref{eq:model:order_parmeter:irreducible_representation}), the resulting order-parameter components $\Dxtyt(\vpF,\vR)$ and $\Dxy(\vpF,\vR)$ are referred to as the nodal components. We initially assume that these channels are degenerate, since such a degeneracy is guaranteed by symmetry in any material with a three- or six-fold rotationally symmetric lattice \cite{Black-Schaffer:2014:b}, relevant for many of the recently proposed chiral $d$-wave superconductors \cite{Venderbos:2018,Su:2018,Fidrysiak:2018,Xu:2018,Kennes:2018,Liu:2018,Gui:2018,Wu:2019,Fischer:2021,Ming:2023,Biswas:2013,Fischer:2014,Ueki:2019,Ueki:2020}. Still, for sake of full completeness, we later relax this assumption. Furthermore, we note that our theoretical framework includes other pair correlations allowed by symmetry, e.g.~$s$-wave \cite{SuperConga:2023}, while the possibility of additional attractive interactions in other pair channels is left as an outlook \cite{Black-Schaffer:2013}.

In order to better quantify chiral superconductivity, we transform the nodal order parameters to the eigenbasis
\begin{align}
    \label{eq:eigenbasis:OAM}
    \etapm(\vpF) \equiv e^{\pm i|M|\thetaF},
\end{align}
of the OAM operator $\Lzorb = (\hbar/i)\partial_{\thetaF}$ with eigenvalues $l^{\mathrm{orb}}_z = \pm|M|\hbar$, yielding
\begin{align}
    \label{eq:chiral_OP}
    \Delta(\vpF,\vR) & = \Dplus(\vpF,\vR) + \Dminus(\vpF,\vR)
\end{align}
with the chiral order parameter components
\begin{align}
    \label{eq:chiral_OP:components}
    \Dpm(\vpF,\vR) & \equiv |\Dpm(\vR)|e^{i\chipm(\vR)}\etapm(\vpF),
\end{align}
which are the two degenerate ground states in a bulk chiral superconductor. Below $\Tc$ the system spontaneously chooses one of these as the dominant bulk chirality, e.g~$\Delta(\vpF,\vR) = \Dplus(\vpF,\vR)$, while the opposite chirality $\Dminus(\vpF,\vR)$ becomes subdominant and vanishes asymptotically in the translationally invariant bulk \cite{Hess_Tokuyasu_Sauls:1989}. Thus, the ground state of a chiral superconductor is described by a complex-valued order parameter that spontaneously breaks time-reversal symmetry \cite{Rice:1995,Volovik:1997,Sigrist:1998}, with a fully gapped bulk spectrum and Cooper pairs with an OAM $l^{\mathrm{orb}}_z = \pm|M|\hbar$ \cite{Sauls:1994}. Even (odd) $|M|$ correspond to spin-singlet (spin-triplet), and $|M|=1,2$ generate chiral $p,d$-wave order parameters, respectively. In this work we focus on spin-singlet chiral $d$-wave superconductivity with $|M|=2$, such that $\eta_\pm(\vpF) = [\etaxtyt(\vpF) \pm i\etaxy(\vpF)]/\sqrt{2}$, which when equating Eq.~(\ref{eq:model:order_parmeter:irreducible_representation}) with Eq.~(\ref{eq:chiral_OP}) yields the relation between the two parametrizations
\begin{align}
    |\Dxtyt(\vR)|e^{i\chixtyt(\vR)} = \frac{1}{\sqrt{2}}\Big(|\Dplus(\vR)|e^{i\chi_+(\vR)} \nonumber \\
    \label{eq:chiral:transform:delta_dx2y2}
    + |\Dminus(\vR)|e^{i\chi_-(\vR)}\Big),\\
    |\Dxy(\vR)|e^{i\chixy(\vR)} = \frac{i}{\sqrt{2}}\Big(|\Dplus(\vR)|e^{i\chi_+(\vR)} \nonumber \\
    \label{eq:chiral:transform:delta_dxy}
    - |\Dminus(\vR)|e^{i\chi_-(\vR)}\Big).
\end{align}

In a chiral superconductor, the topological invariant is the Chern number $M$ corresponding to the winding of the superconducting order parameter on the FS and giving rise to $|M|$ chiral edge modes traversing the bulk gap whenever the topoloigcal invariant changes, in particular at vacuum interfaces but also domain walls \cite{Volovik:1997,Schnyder:2008,Hasan:2010,Qi:2011,Sauls:2011,Tanaka:2012,Graf:2013,Black-Schaffer:2014:b}. While these edge modes are topologically protected, they generate chiral edge currents and OAM which are not \cite{Volovik:1988,Black-Schaffer:2012,Nie:2020}. Furthermore, close to the edges, the opposite (subdominant) chirality is often also locally induced, such that the order parameter takes the more general form in Eq.~(\ref{eq:chiral_OP}). This extends more generally to other forms of spatial inhomogeneities such as domain walls and vortices, and we therefore always use the most general form in Eq.~(\ref{eq:chiral_OP}) in our calculations, allowing for a completely general spatial dependence of both amplitudes and phases. We note that this in principle allows the system to go into a different state, e.g.~a nodal $d$-wave or nematic $d$-wave state \cite{Lothman:2022}, but we always find the chiral state to be robust.

Chiral superconductors also hosts domain walls, which are topological defects separating regions of opposite dominant chirality \cite{Volovik:2003,Beri:2010,Becerra:2016:a}. Domain walls thus have $|M|$ chiral edge modes on each side with opposite winding \cite{Awoga:2017}, also generating chiral currents on either side. These currents, together with the exchange of chirality across the domain wall, lead to a slight increase in free energy and an effective line tension \cite{Garaud:2013}. This usually makes domain walls metastable, but they are often trapped and further stabilized by pinning, geometric effects, and vortices \cite{Garaud:2014}.

Just like any superconductor, a chiral superconductor can also host vortex defects.
A chiral superconductor with total vorticity $m$ is associated with an $m\times2\pi$ quantized phase winding in the dominant chiral component \cite{Sauls:2009}, i.e.~$\chi_+(\vR) \approx m\phi$ along any path sufficiently far from and encircling all vortex defects. Abrikosov vortices (antivortices \cite{Chibotaru:2000,Zhang:2012,Zhang:2013,Iavarone:2011,Giorgio:2017,Simmendinger:2020}) correspond to $m=-1$ ($m=+1$) in positive external flux $\Phiext>0$, and vice versa for negative flux, also with a corresponding $2\pi$ phase winding in each nodal component $\chixtyt(\vR)$ and $\chixy(\vR^\prime)$ if the vortex cores are overlapping, $\vR=\vR^\prime$. Spatially separating the nodal winding centres $\vR\neq\vR^\prime$ leads to a disassociation of the Abrikosov vortex into two fractional vortices, one for each winding center, and to a Josephson-like term in the free energy that usually grows with the separation distance \cite{Sigrist:1989,Sigrist:1999,Etter:2020}, thus making the fractional vortices unstable. However, inside a domain wall such a separation typically becomes favorable instead \cite{Garaud:2013}. Furthermore, the slight suppression of the total order parameter in the domain wall acts as an attractive pinning center for Abrikosov vortices, providing a mutual stabilization of the domain wall and fractional vortices \cite{Sigrist:1990}, and thereby a mechanism for forming a CV as demonstrated in the next section \ref{sec:coreless_vortices:basics}.

Finally, changing magnetic flux direction allows for the vorticity to either be aligned antiparallel or parallel with the chirality, which leads to inequivalent vortices and also to inequivalent CVs. We illustrate this by first considering the total angular momentum, $\Lz = \Lzorb + \Lzcm$, and the winding quantization. Here, $\Lzorb$ is the OAM generated by chirality as explained earlier in this subsection, while $\Lzcm = (\hbar/i)\partial_{\phi}$ is the generator of c.m.~angular momentum with eigenvalue $l^{\mathrm{c.m.}}_z = m\hbar$ for a state with vorticity $m$. Thus, the total angular momentum of the Cooper pair is $l_z = l^{\mathrm{orb}}_z + l^{\mathrm{c.m.}}_z = (M+m)\hbar$, and is therefore a superposition between the OAM generated by chirality (i.e.~Chern number) and the c.m. angular momentum generated by vorticity (i.e.~winding quantization). Thus, antiparallel (parallel) alignment of vorticity and chirality leads to a negative (positive) superposition of the total angular momentum. Similarly, the phase winding of the subdominant chirality also shows such a behavior. Close to a vortex defect, the subdominant chirality is generally induced with finite amplitude and phase $\chi_-(\vR) \approx p\phi$. The quantized phase winding $p$ is constrained according to the relation \cite{Sauls:2009,Holmvall:2023:arxiv}
\begin{align}
    \label{eq:phase_constraint}
    p = m + 2M + n,
\end{align}
here with integer $n$ capturing higher-order harmonics generated by e.g.~a non-circular system or anistropic FS \cite{Yip:1993}. Despite such terms often being unimportant \cite{Sauls:2009}, we in this work include them for full completeness. Equation~(\ref{eq:phase_constraint}) shows that the phase winding of the subdominant component also is a superposition of the vorticity and Chern number and can therefore be minimized (maximized) for an antiparallel (parallel) alignment. 

\section{Coreless vortices}
\label{sec:coreless_vortices}
We begin this section by briefly summarizing the basic structure and properties of CVs in spin-singlet chiral $d$-wave superconductors in Sec.~\ref{sec:coreless_vortices:basics}. In Sec.~\ref{sec:coreless_vortices:stability} we discuss the stability and formation of CVs, and that the most stable CV is typically quadruply quantized in chiral $d$-wave superconductors.

\subsection{Coreless vortex structure}
\label{sec:coreless_vortices:basics}
In this subsection we summarize the basic properties of antiparallel and parallel CVs. In comparison to our earlier work \cite{Holmvall:2023:arxiv}, we here choose to study a somewhat smaller system with slightly different parameters, to illustrate that the important qualitative features do not depend on such parameters. Note that the spatial inhomogeneities induced by the CVs therefore are significant compared to the system size. Still, when we use the term `dominant bulk chirality', we refer to the spontaneously chosen ground state chirality in the absence of vorticity, or equivalently, the dominant chirality in a much larger but otherwise analogous system. For reference, see Appendix~\ref{app:larger_systems} showing that the important qualitative features discussed here remain in systems with radius $\mathcal{R}\geq150\xi_0$, i.e.~an order of magnitude larger.

\begin{figure*}[tb!]
    \centering
    \begin{minipage}[t]{0.49\textwidth}
	\includegraphics[width=\columnwidth]{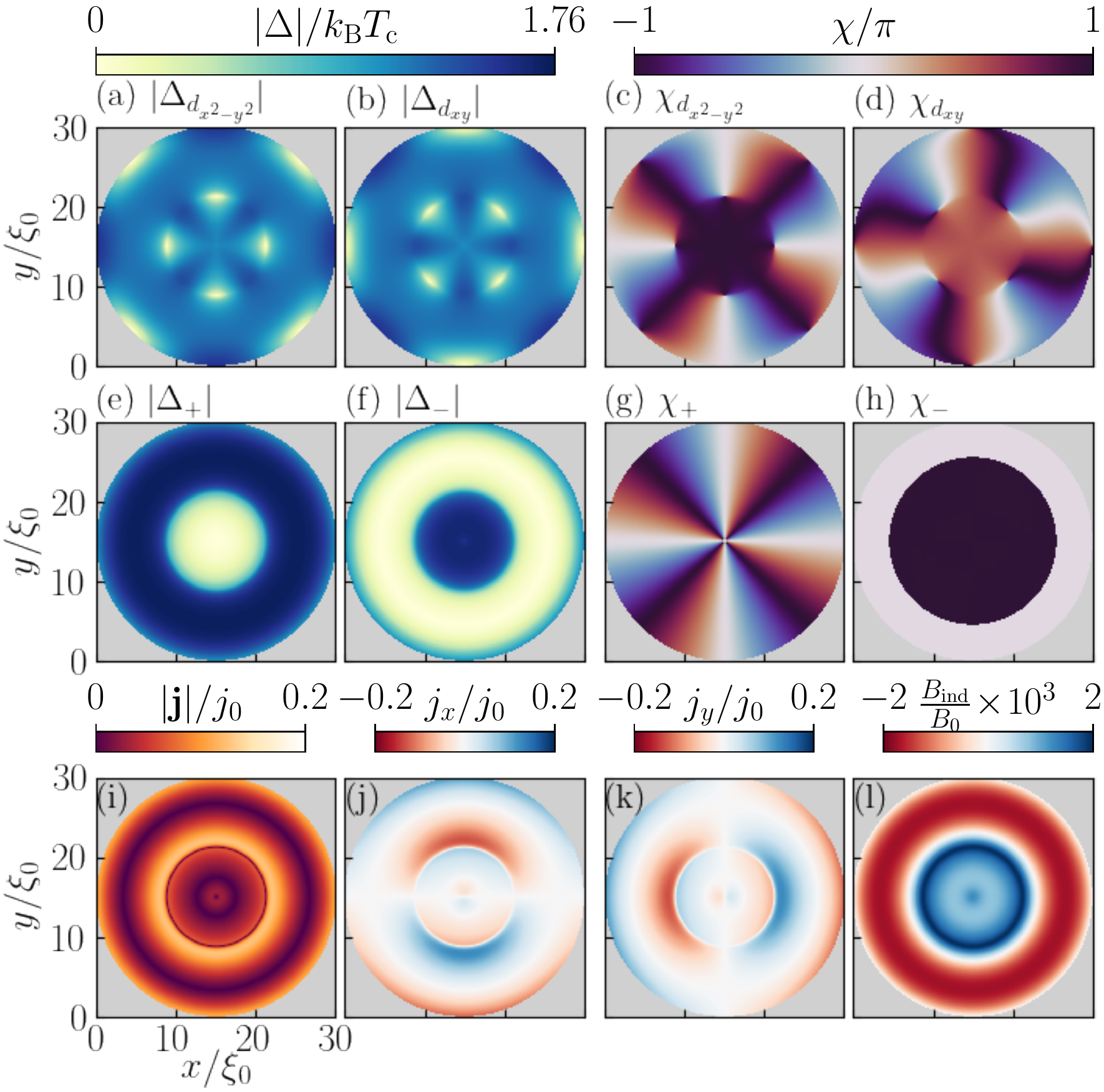}
	\caption{Antiparallel CV in disc-shaped system with radius $\mathcal{R}=15\xi_0$, bulk chirality $\Dplus$, $T=0.1\Tc$, $\lambda_0=10\xi_0$, $\Phiext=7.5\Phi_0$. First (second) row: nodal (chiral) component amplitudes and phases. Third row: magnitude and $x,y$-components of charge-current density $\vj$, and induced magnetic-flux density $\Bind$. Natural units: $j_0 \equiv \hbar|e|\vF^2\NF/\xi_0$, $B_0 \equiv \Phi_0/(\pi\xi_0^2)$, $\Phi_0\equiv hc/2|e|$.
     }
	\label{fig:antiparallel_CV}
    \end{minipage}\hfill
    \begin{minipage}[t]{0.49\textwidth}
	\includegraphics[width=\columnwidth]{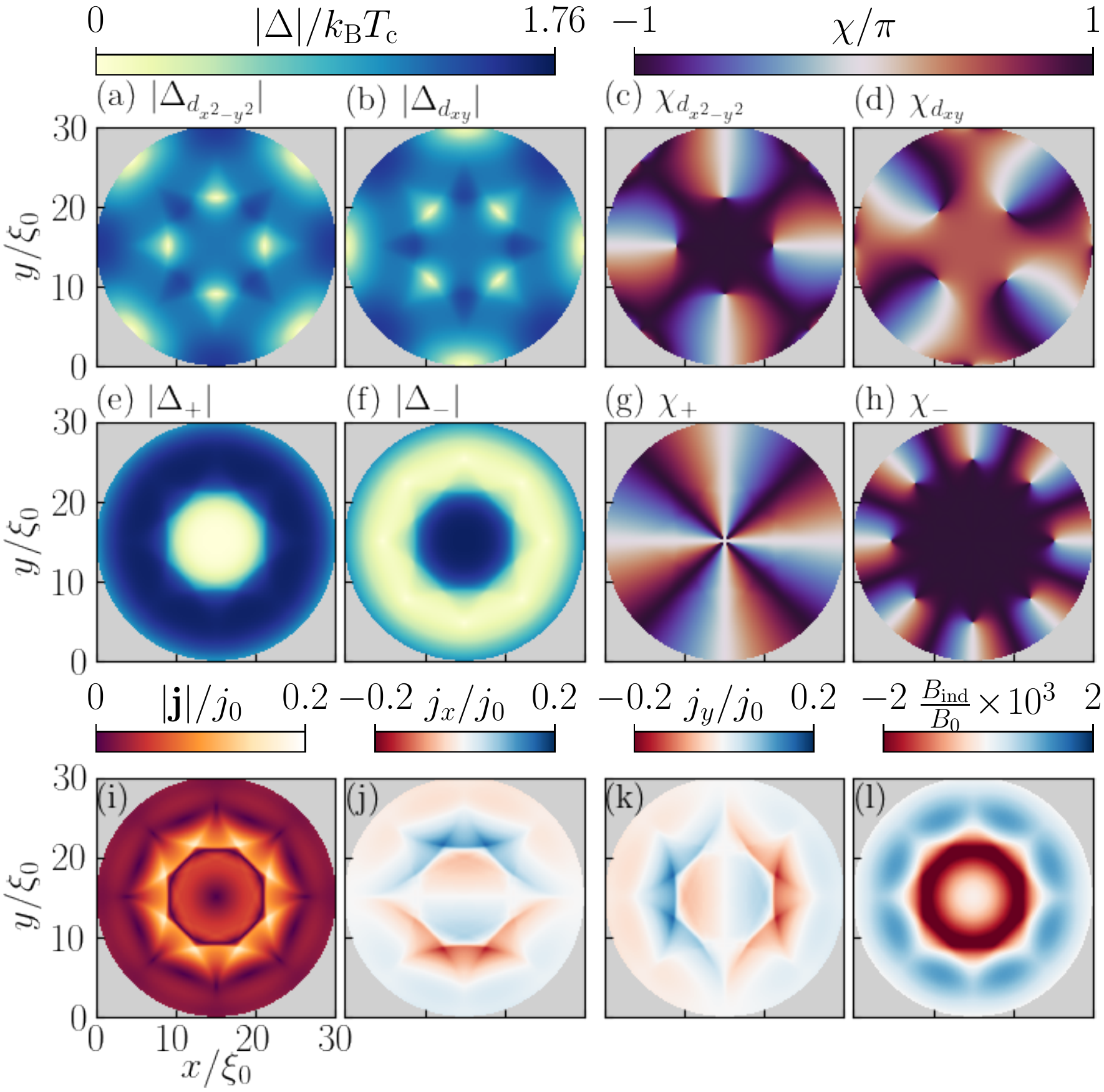}
	\caption{Same as Fig.~\ref{fig:antiparallel_CV} but for a parallel CV with $\Phiext=-7.5\Phi_0$.
     }
	\label{fig:parallel_CV}
    \end{minipage}
\end{figure*}

Figure~\ref{fig:antiparallel_CV} (Fig.~\ref{fig:parallel_CV}) shows an antiparallel (parallel) CV in a disc-shaped system with radius $\mathcal{R}=15\xi_0$, dominant bulk chirality $\Dplus$, temperature $T=0.1\Tc$, penetration depth $\lambda_0=10\xi_0$, and external flux $\Phiext=7.5\Phi_0$ ($\Phiext=-7.5\Phi_0$). The first (second) row shows the amplitudes and phases of the nodal (chiral) components, while the third row shows the charge-current density and induced magnetic-flux density. Each nodal component has four $2\pi$-phase windings that suppress the corresponding nodal amplitude but not the other nodal component. Since they lie at different coordinates for the two nodal components, the total order parameter is everywhere non-singular with no normal-state regions or core. The vortices are therefore fractional, in contrast to singular Abrikosov vortices which have spatially overlapping phase windings. A total of eight fractional vortices lie on a circularly (octagonally) formed domain wall, the latter seen in the second row of Fig.~\ref{fig:antiparallel_CV} (Fig.~\ref{fig:parallel_CV}), where it separates the outer and inner regions of dominant chiralities $\Dplus$ and $\Dminus$, respectively. There is a total vorticity of $m=\pm4$ in the disc, seen by the $m\times2\pi$ winding of the dominant chirality $\chi_+(\vR) = m\phi$ in the outer region, which means this is a quadruply quantized CV. There is no phase-winding in the inner region, since there the dominant phase is constant $\chi_-=0$, indicating that the vorticity is distributed along the domain wall.
We next turn to the subdominant phase, which shows a $\pi$-shift across the domain wall in Fig.~\ref{fig:antiparallel_CV}, which further stabilizes the structure but is otherwise unimportant \cite{Holmvall:2023:arxiv}. Thus, apart from this phase shift, the phase $\chi_-$ is completely trivial in Fig.~\ref{fig:antiparallel_CV}. In contrast, Fig.~\ref{fig:parallel_CV} shows a total of eight winding centers in the subdominant component $\chi_-$ in the outside region. These results are in full agreement with the phase winding constraint in Eq.~(\ref{eq:phase_constraint}), with $p=-4+4=0$ ($p=4+4=8$) in Fig.~\ref{fig:antiparallel_CV} (Fig.~\ref{fig:parallel_CV}) corresponding to a CV with antiparallel (parallel) alignment of vorticity and chirality. Importantly, Fig.~\ref{fig:parallel_CV} shows that the winding centers lie outside the CV, spontaneously breaking axial symmetry, as defined by the winding not being generated by rotation around a single central axis. This occurs in order to lower the free energy, since the hypothetical axisymmetric state with $p=8$ would correspond to a giant vortex with a large normal core \cite{Sauls:2009}, consequently suppressing superconductivity and increasing the free energy. We have verified that such a giant vortex is indeed unstable. In contrast, the axially symmetry-breaking CV is stable since it avoids the energy penalty, while still lowering the kinetic energy caused by the external flux. Thus, the antiparallel (parallel) CV is axisymmetric (non-axisymmetric) with a continuous (discrete eight-fold) rotational symmetry. Our earlier work showed that this leads to a smoking-gun signature in the LDOS of both the topologically protected and quantized Chern number and vorticity \cite{Holmvall:2023:arxiv}. The third row shows a corresponding rotation symmetry of the charge-current density and induced magnetic flux, with multiple sign changes due to the chiral edge modes, domain wall, and overlapping Meissner screening currents. The paramagnetism is maximal along the domain wall, leading to a characteristic ring-like magnetic structure, in contrast to a point-like structure of an Abrikosov vortex \cite{Holmvall:2023:arxiv}.

Overall these results demonstrate the structure and basic properties of quadruply quantized CVs in chiral $d$-wave superconductors, i.e.~quadruple-quantum vortices. This is the chiral $d$-wave extension of the double-quantum vortex in chiral $p$-wave superfluids \cite{Sauls:2009,Garaud:2012,Takamatsu:2013,Garaud:2015,Garaud:2016,Zhang:2016,Becerra:2016:b,Zyuzin:2017,Zha:2020,Chai:2021,Krohg:2021}. Beyond this comparison, we note that an extension between the two different systems can in general be very non-trivial due to the different spin-symmetries and angular momentum quantization, and therefore it is not \textit{a priori} certain that the same kind of vortex defects are even stable in both systems, let alone have the same qualitative properties. For example, the parallel CV in Fig.~\ref{fig:parallel_CV} shows multiple sign changes, compared to no sign changes for the parallel CV in a chiral $p$-wave double-quantum vortex reported in Ref.~\cite{Sauls:2009}. More generally, ``$p$-wave is special'' \cite{Huang:2014} in many regards compared to all systems with higher Chern number, e.g.~when it comes to the chiral edge currents and OAM \cite{Huang:2014,Tada:2015,Volovik:2015:b,Suzuki:2016,Ojanen:2016,Wang:2018,Tada:2018,Nie:2020,Sugiyama:2020}.

\subsection{Stability and coreless vortex formation}
\label{sec:coreless_vortices:stability}
We next discuss the stability and formation of CVs. The peculiar combination of a domain wall and vorticity in a CV allows the system to carry finite vorticity, which reduces the kinetic energy caused by external flux but without paying the price of a normal core. This is significant, since the superconducting state is per definition the most energetically favorable state below the second critical field $\BcTwo(T)$. The CV will thus be energetically more favorable than Abrikosov vortices if this gain outweighs the cost of the domain wall. However, the CV is very robust even when there are other vortex configurations with a lower free energy, i.e.~even when technically metastable. This is a general feature of both Abrikosov vortices and CVs, related to the fact that they are topological defects that cannot be trivially removed from the system. They typically have to enter and exit the system via the edges, but such entrance and expulsion is hampered by large energy barriers, e.g.~geometric and Bean-Livingston barriers \cite{Bean:1964, Burlachkov:1991, Volovik:2015:a, Benfenati:2020:b, Maiani:2022, SuperConga:2023}. Moreover, vortex motion is hampered by pinning and dissipation associated with normal-state resistance. Thus, once a particular arrangement of vortex defects has entered the system, it can become extremely robust even far into the flux-temperature regime where other vortex arrangements technically have even a significantly lower energy. Summarized briefly, experiments to a large degree observe metastable states \cite{Blatter:1994}, and such behavior is also typical in self-consistency simulations. Thus, the most relevant question is not necessarily whether a particular vortex configuration has the lowest energy, but if the necessary conditions for its formation can be prepared \cite{Volovik:2015:a}.

We see the vortex stability repeatedly in our simulations, both for CVs and Abrikosov vortices. In particular, we find that CVs spontaneously enter the system instead of Abrikosov vortices in certain parameter regimes, or can easily form when both domain walls and vortices are present. In the latter scenario, we find that the domain wall attracts and pins the Abrikosov vortices and, upon entering the domain wall, they disassociate into fractional vortices that lowers the free energy \cite{Sigrist:1989,Sigrist:1999,Garaud:2012,Garaud:2013,Etter:2020}. Conversely, to break the CV, the vortices have to exit the domain wall or the domain wall has to disappear. However, such vortex expulsion is prevented by the pinning, and more importantly, the instability of the fractional vortices outside the domain wall. Thus, fractional vortices typically first have to re-combine to a regular Abrikosov vortex before expulsion, but such re-combination increases the free energy. For the domain wall to disappear, it either has to shrink to zero size or expand to the system edges. Such shrinking is however prevented by the strong repulsive interaction between vortices, while expansion of the CV is counter-balanced by the attractive interaction (line tension) caused by the domain-wall currents, as well as the repulsive interaction between the fractional vortices and system boundaries. Among the very rare instances where we find the CV becoming unstable, it is this latter scenario that seems the most plausible; the line tension is significantly modified by non-degenerate order-parameter components (e.g.~competing nodal superconductivity discussed in Sec.~\ref{sec:non_degeneracy}) or the CV expanding to the system edge combined with a flux-temperature combination very far from the energy minimum (e.g.~in Sec.~\ref{sec:cv_size}). Apart from these scenarios, we find the CV to be extremely robust in all our calculations and often spontaneously appearing, even in the presence of strong perturbations, disorder, and when there are other vortex configurations with considerably lower free energy.

Finally, we discuss the most stable CV, which we generally find to be the quadruply quantized CV with $|m|=4$, shown in Figs.~\ref{fig:antiparallel_CV} and \ref{fig:parallel_CV}, and discussed in our previous work \cite{Holmvall:2023:arxiv}. This is easy to understand for the antiparallel CV, since it corresponds to the special commensurate scenario, such that the phase winding of the subdominant chirality vanishes $p=m+2M=0$ [Eq.~(\ref{eq:phase_constraint})]. In contrast, a finite phase winding $p>0$ would either suppresses superconductivity if axisymmetric (thus costing energy), or increase the phase winding generating a modified superfluid momentum and line tension if non-axisymmetric (also costing energy). Furthermore, beyond commensurability, there is also the matter of balancing the repulsive versus attractive interactions, which overall stabilize the CV and its finite size (as discussed in Sec.~\ref{sec:cv_size}), which is then important for the parallel CV, since there cancellation in $p$ is impossible by definition. Considering for example higher vorticity $|m|>2|M|$, this leads to increased repulsion, but also modified line tension due to the additional phase windings in $p$. As a consequence, the CV becomes less energetically favorable and less robust. The latter is also true for lower vorticity $|m| < 2|M|$, as there here might no longer be enough vortices to stabilize the domain wall. We verify these arguments during the extensive self-consistency calculations of the present work, including the large parameter ranges and model comparisons. Although we have found some parameter regimes where CVs with higher or lower vorticity become metastable rather than completely unstable, these states were generally less favorable and significantly more difficult to get to appear in the system. In summary, the commensurate scenario $m=-2M$ allows the antiparallel CV to be coreless with maximized order parameter, leading to quadruple-quantum vortices in chiral $d$-wave superconductors, and more generally $2|M|$-quantum vortices for other chiral superfluids.

\section{Tunable coreless vortex size}
\label{sec:cv_size}
This section demonstrates the large tunability of the CV size, via easily accessible parameters in experiment such as external flux $\Phiext$ and temperature $T$, but also via the penetration depth $\lambda_0$ and system size $\mathcal{R}$. For all the parameter ranges considered in this section, we note that the overall qualitative features presented in Figs.~\ref{fig:antiparallel_CV} and \ref{fig:parallel_CV} remain the same.

\begin{figure}[tb!]
	\includegraphics[width=\columnwidth]{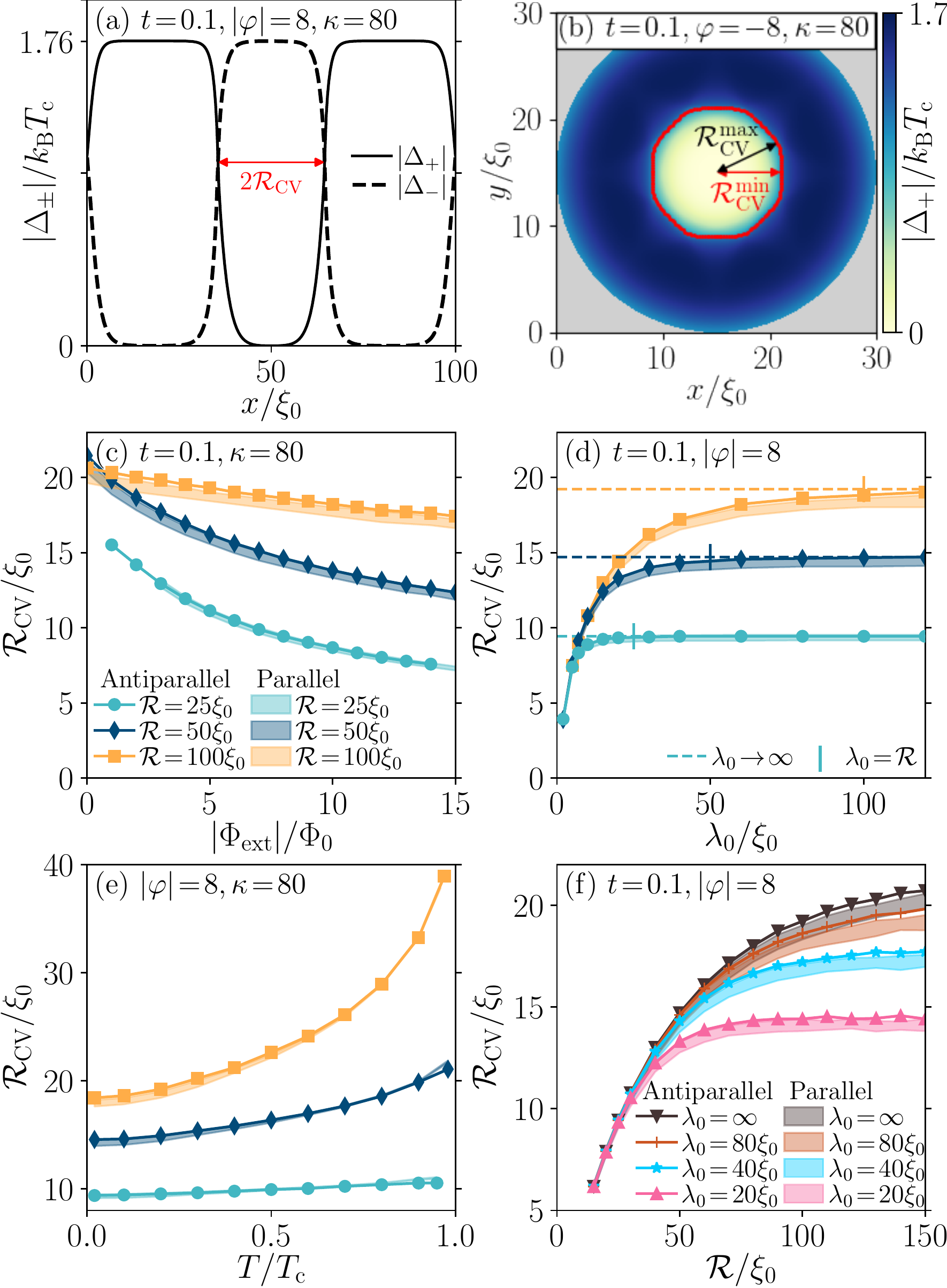}
	\caption{Tunability of the CV radius $\RCV$. (a) Line cut through the center of a CV, defining $\RCV$ as half the distance between the two degeneracy points $|\Dplus| = |\Dminus|$. Here, $t \equiv T/\Tc$, $\varphi \equiv \Phiext/\Phi_0$, and $\kappa=\lambda_0/\xi_0$. (b) Heatmap of $|\Dplus|$ for parallel CV, showing how $\RCV$ varies with angle, with red line indicating $|\Dplus| = |\Dminus|$ and minimum (maximum) CV radius $\RCV^{\mathrm{min}}$ ($\RCV^{\mathrm{max}}$). $\RCV$ versus external flux (c), penetration depth (d), temperature (e), system radius (f) for antiparallel CV (markers) and parallel CV (shaded regions $\RCV^{\mathrm{min}}$ and $\RCV^{\mathrm{max}}$).
     }
	\label{fig:coreless_vortex:radius_comparison}
\end{figure}

The CV has a finite radius, $\RCV$, balanced by attractive and repulsive interactions, acting to contract and expand the CV, respectively \cite{Volovik:2015:a}. The attractive interaction is exerted by the effective line tension from the domain wall and its chiral currents \cite{Garaud:2013}, while there is a mutual repulsive interaction between the fractional vortices in the domain wall \cite{Sigrist:1989,Sigrist:1999}. Hence, a closed domain wall will typically collapse and disappear in the absence of vorticity (we have verified this in our self-consistent calculations) \cite{Garaud:2013}. Furthermore, anything influencing the currents or vortices will change the balance, and therefore also $\RCV$. This is also further demonstrated by studying the interaction between CVs and Abrikosov vortices in Sec.~\ref{sec:cv_v_interaction} or with the system edges in Sec.~\ref{sec:non_circular_grains}.

We start by describing how to unambiguously define and calculate $\RCV$ for antiparallel and parallel CVs. The midpoint of the CV is always well-defined, and a straight line across this point will generally intersect the domain wall of the CV twice, i.e.~in two different points with degeneracy $|\Dplus|=|\Dminus|$ as indicated in Fig.~\ref{fig:coreless_vortex:radius_comparison}(a). We note that these points generally coincide with the maximum of the zero-energy LDOS \cite{Holmvall:2023:arxiv}. For the antiparallel CV, the CV diameter is the distance between the intersection points and is independent of the angle, and $\RCV$ is therefore unambiguously defined as half this distance as in Fig.~\ref{fig:coreless_vortex:radius_comparison}(a). For the parallel CV, we instead define $\RCV$ from the average half distance for all angles, as displayed in Figs.~\ref{fig:coreless_vortex:radius_comparison}(b) where the thick red line shows the numerically extracted point $|\Dplus|=|\Dminus|$ and arrows show the minimum (maximum) radius $\RCV^{\mathrm{min}}$ ($\RCV^{\mathrm{max}}$). However, we find that $\RCV$ is practically unambiguously defined even for the parallel CV, since $\Delta\RCV \equiv \RCV^{\mathrm{max}}-\RCV^{\mathrm{min}} \lesssim 1\xi_0$ in all our simulations across all parameter ranges.

In Fig.~\ref{fig:coreless_vortex:radius_comparison}(c) we illustrate that $\RCV$ can be effectively tuned by an externally applied magnetic field. Specifically, $\RCV$ decreases as $|\Phiext|$ increases, since the currents grow in magnitude, while the distance between fractional vortices reduce (hence an overall stronger contraction). This is in a sense analogous with how larger flux causes smaller vortex separation and denser vortex lattices in regular type-II superconductors \cite{Tinkham:2004}. Similarly, a shorter penetration depth $\lambda_0$ also leads to a smaller vortex-vortex separation, implying a smaller effective repulsive interaction as seen in Fig.~\ref{fig:coreless_vortex:radius_comparison}(d). The overall dependence on $\lambda_0$ can be divided into two regimes: $\lambda_0 < \mathcal{R}$, where screening becomes considerable and strongly modifies $\RCV$, and $\lambda_0 > \mathcal{R}$, where the system is poorly screened and the effect is minimal. In the limit of small $\lambda_0$ such that $\xi_0 \lesssim \lambda_0 \ll \mathcal{R}$, the CV radius is almost completely determined by the screening regardless of system size, while in the opposite limit of large $\lambda_0$, $\RCV$ eventually reaches the asymptotic limit $\lambda_0 \rightarrow \infty$ (zero screening). We here note that the penetration depth is a materials property, which can be modified by the inclusion of impurities, as non-magnetic and magnetic impurities typically increase and decrease the penetration depth, respectively \cite{Prozorov:2022}. 

Figure~\ref{fig:coreless_vortex:radius_comparison}(e) shows that $\RCV$ decreases at lower temperatures. We interpret the overall temperature dependence to be directly proportional to the effective coherence length ${\xieff\equiv\hbar\vF/|\Delta(T)|}$, which reduces but saturates at small temperatures (due to saturating $|\Delta(T)|$) and increases dramatically at large temperature (due to vanishing $|\Delta(T)|$). Note that this is consistent with stronger but saturating chiral currents at lower temperatures, hence increasing the contraction. Of course, $\RCV$ is strictly limited by the size of the system (relative to the coherence length), consistent with the observed small (large) temperature dependence in small (large) systems in Fig.~\ref{fig:coreless_vortex:radius_comparison}(e). Specifically, in the small systems there is a strong overlap between the boundary and CV at all temperatures leading to saturation, while the much weaker overlap in larger systems leave considerably more room for variation in $\RCV$ with $T$. We note an overall trend that $\RCV \to 0.4 \mathcal{R}$ for large temperature, for all system sizes $\mathcal{R}$ considered in our simulations. In other words, the system size, and more generally surrounding environment, can strongly influence the maximum $\RCV$ and its temperature dependence.

Figure~\ref{fig:coreless_vortex:radius_comparison}(f) shows directly how $\RCV$ increases with system size $\mathcal{R}$. This is a mesoscopic finite-size effect, which can be divided into two regimes, corresponding to small and large $\mathcal{R}$. For small $\mathcal{R}$, the CV-induced currents strongly overlap with the chiral edge currents of the system. More importantly, the system edges impose an energy barrier \cite{Bean:1964, Burlachkov:1991, Volovik:2015:a, Benfenati:2020:b, Maiani:2022, SuperConga:2023} and an effective repulsive interaction (at least at sufficiently high flux), which contracts the CV. This effect is also seen in Sec.~\ref{sec:non_circular_grains}, and is well-known for vortex lattices, leading to a number of interesting mesoscopic finite-size and shape effects \cite{Geim:1997,Grigorieva:2006,Zhao:2008_a,Zhao:2008_b,Misko:2009,Cabral:2009,Kokubo:2010,Zhang:2012,Zhang:2013,Huy:2013,Timmermans:2016,Kokubo:2016,Wu:2017}, see also Ref.~\cite{SuperConga:2023} and references therein. For large $\mathcal{R}$, the repulsion from the edges eventually becomes negligible, but there is still a slow asymptotic behavior of $\RCV$ which we interpret to be due to a slow saturation also present in properties related to the spectrum and chiral currents surrounding the CV. 

In summary, these results demonstrate a strong tunability of the CV size, traced back to the effective attractive and repulsive interactions balancing the finite size \cite{Volovik:2015:a}, but also to the effective coherence length and its dependence on the superconducting gap. We note that while there are significant differences in the LDOS for the antiparallel and parallel CVs, due to symmetry breaking for the latter, the overall CV size is roughly simlar for both CVs.
In the next Sec.~\ref{sec:cv_v_interaction} and Sec.~\ref{sec:non_circular_grains} we further demonstrate a tunability of the CV shape in the presence of other vortices or anisotropic effects.

\section{Interaction with Abrikosov vortices}
\label{sec:cv_v_interaction}
\begin{figure*}[bt!]
	\includegraphics[width=\textwidth]{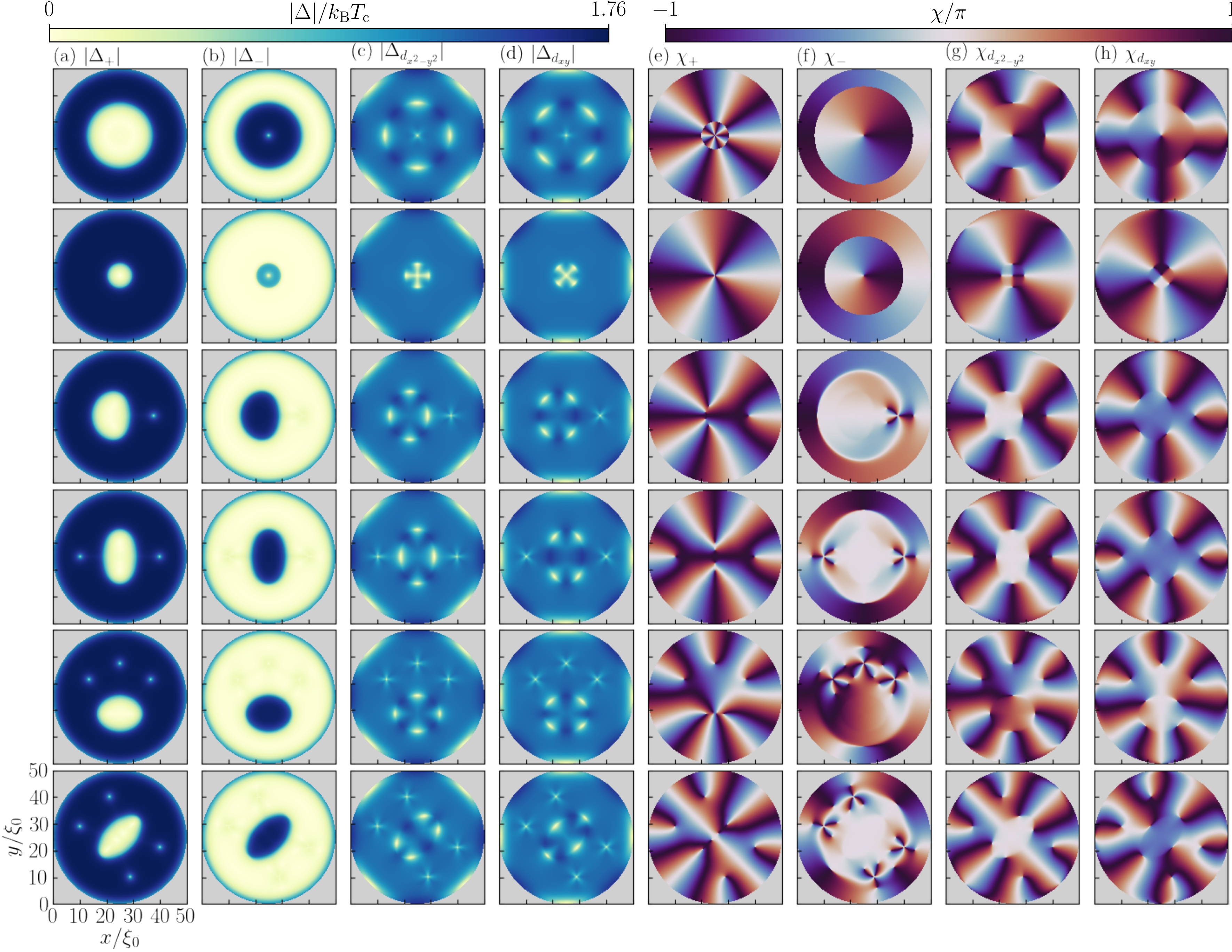}
	\caption{Systems containing both an antiparallel CV and Abrikosov vortices or antivortices, in a disc-shaped system with radius $\mathcal{R}=25\xi_0$, dominant bulk chirality $\Dplus$, at $T=0.1\Tc$ and $\lambda_0=80\xi_0$. Columns show, from left to right, the chiral and nodal amplitudes, then the chiral and nodal phases. In the first (second) row, an Abrikosov (anti)vortex trapped inside the CV, remaining rows an increasing number of Abrikosov vortices (one to four) outside the CV. External flux is from top to bottom row: $8\Phi_0$, $8\Phi_0$, $12\Phi_0$, $12\Phi_0$, $14\Phi_0$, and $14\Phi_0$ in order to stabilize the various configurations. }
	\label{fig:cv_v:quantities}
\end{figure*}

\begin{figure*}[bt!]
	\includegraphics[width=\textwidth]{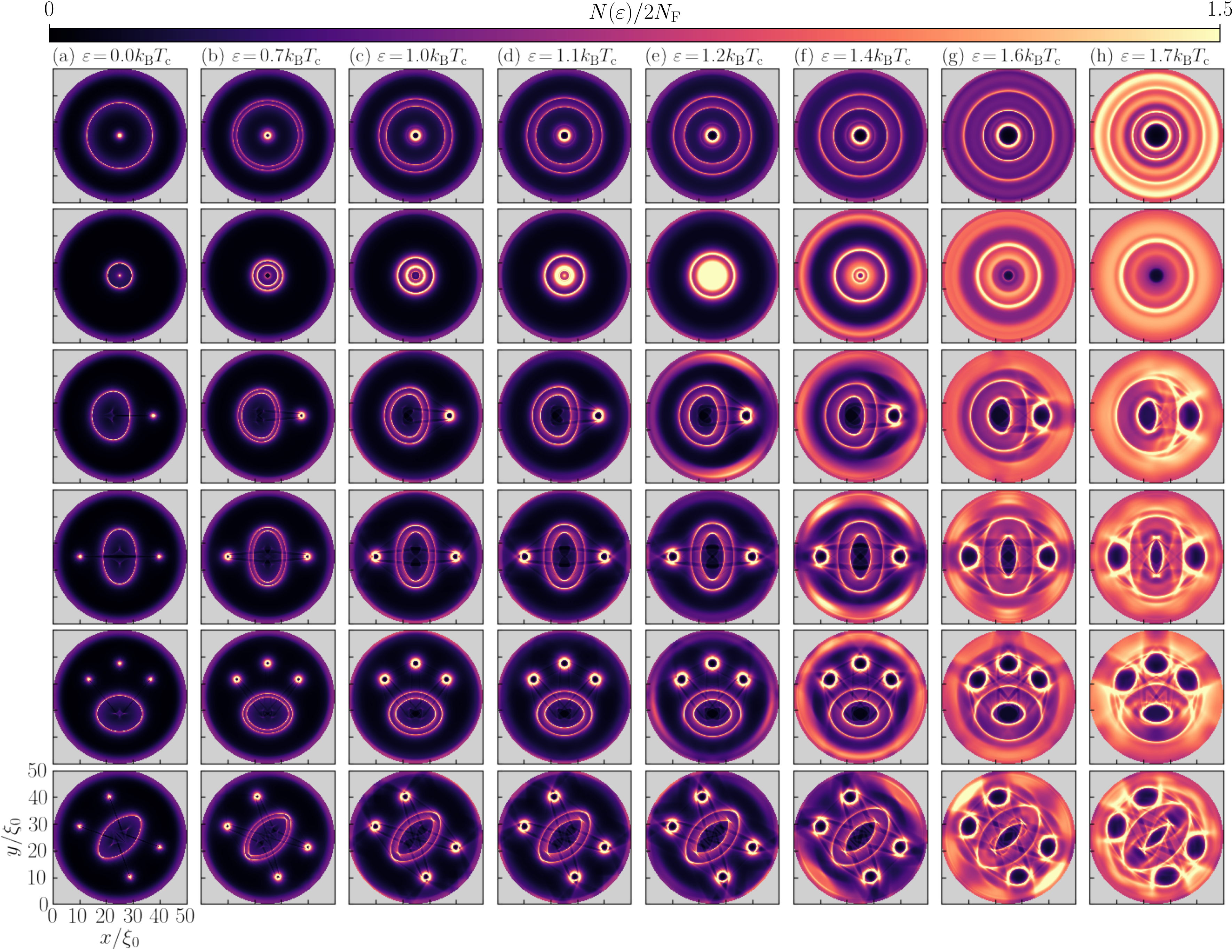}
	\caption{Same as Fig.~\ref{fig:cv_v:quantities}, but with each column showing the LDOS at a fixed subgap energy $\varepsilon$, with gap roughly $1.76\kB\Tc$.}
	\label{fig:cv_v:LDOS}
\end{figure*}

\begin{figure*}[bt!]
	\includegraphics[width=\textwidth]{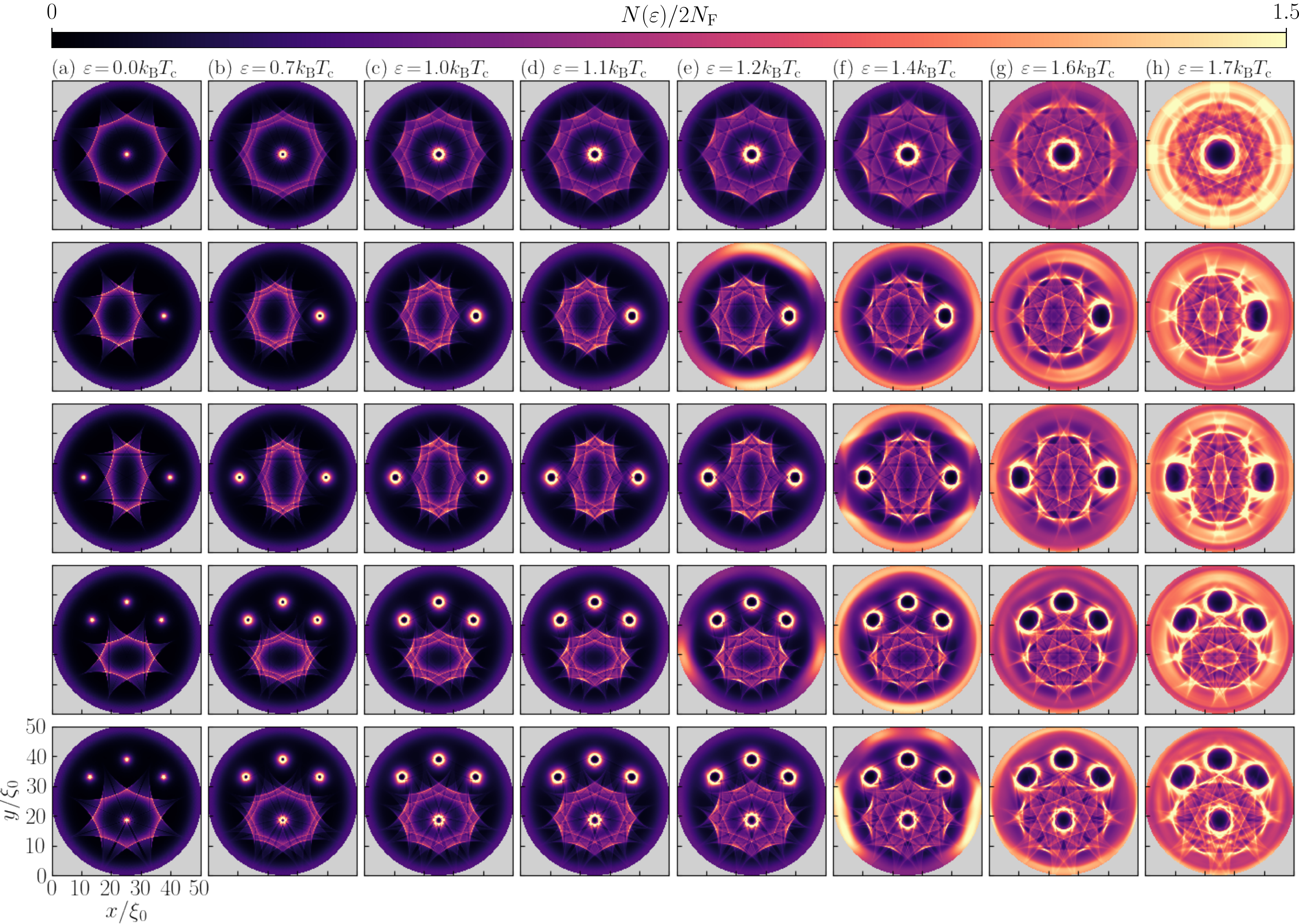}
	\caption{Same as Fig.~\ref{fig:cv_v:LDOS}, but for a parallel CV, without any antivortex scenario, and where one of the four vortices in the last row has spontaneously been trapped at the CV center. The system has dominant bulk chirality $\Dplus$, at $T=0.1\Tc$ and $\lambda_0=80\xi_0$, and is exposed to negative external fluxes, from top to bottom: $-8\Phi_0$, $-12\Phi_0$, $-12\Phi_0$, $-14\Phi_0$, and $-14\Phi_0$.}
	\label{fig:cv_v:ASB:LDOS}
\end{figure*}

In this section, we address how CVs coexisting with Abrikosov vortices changes the CV shape. In Figs~\ref{fig:cv_v:quantities} and \ref{fig:cv_v:LDOS} we show results for an antiparallel CV, and in Fig.~\ref{fig:cv_v:ASB:LDOS} for a parallel CV.
Before describing these figures in detail, we note that unlike the mostly point-like Abrikosov vortex, the CV has an intrinsic structure whose shape is to a large degree set by the repulsive interaction between its fractional vortices and their interaction with the environment, as established in the previous section \ref{sec:cv_size}. For example, the CV interacts repulsively with other vortices, whether it is Abrikosov vortices or other CVs, or attractively with antivortices. This section also establishes robustness of both the CV and its distinctive LDOS signature in the presence of such strong perturbations as additional vortices, and furthermore also shows the distinctly different LDOS signatures of CVs versus Abrikosov vortices. We note that the section also essentially studies the interaction between fractional vortices and regular Abrikosov vortices. Apart from illustrating all these aspects, combinations of CVs and Abrikosov vortices are reasonable to expect in a chiral $d$-wave superconductor, as discussed in the end of the section.

In Fig.~\ref{fig:cv_v:quantities} we present in the different columns the order parameter amplitudes and phases, while each row represents a different configuration of one antiparallel CV with one or more Abrikosov vortices or an antivortex. We start by analyzing the amplitudes and the overall CV shape, and then analyze the phases.
The figure explicitly shows how the Abrikosov vortices completely suppress both the nodal components $|\Dxtyt|$ and $|\Dxy|$ at its core, as well as both the chiral components $|\Dplus|$ and $|\Dminus|$. By contrast, the fractional vortices in the domain wall of the CV only suppress the corresponding nodal component. 
Importantly, the figure shows significant modification of the overall CV size and shape. To explain these results, we note that at a sufficiently high external flux, Abrikosov vortices and CVs are both repelled from the system edges, related to the geometric barriers and the Bean-Livingston barrier \cite{Bean:1964,Burlachkov:1991,Volovik:2015:a,Benfenati:2020:b,Maiani:2022, SuperConga:2023}. This confinement leads to relatively small distances between Abrikosov vortices and the CV, which deforms the CV due to the mutual repulsive interaction. The resulting shape depends on the exact number and spatial arrangement of the Abrikosov vortices. Hence, we find that different deformation modes appear, as clearly seen in rows three to six. However, we note that the CV still keeps an overall elliptical form, which is clearly traced back to its original unperturbed circular form. If the Abrikosov vortex is instead situated at the center of the CV (first row), it is trapped and the CV expands due to the mutual repulsive interaction. If instead an antivortex is situated inside the CV (second row), it attracts the CV, which then shrinks substantially. However, we find that this configuration is always unstable unless pinning centers are artificially added (thus stabilizing the configuration), since the slightest deviation will otherwise fully attract the antivortex into the CV domain wall where it will be annihilated against two of the fractional vortices. We note that all other results and scenarios considered here are very robust even without such pinning, and we only choose to plot the antivortex scenario as it clearly illustrates how competing attractive and repulsive interactions set the overall shape and size of the coreless vortex. Specifically, all other results show a fully converged self-consistent solution, stabilized and trapped in the system by large energy barriers, and corresponding to a minimum of free energy.

Next, we study the phases and note in particular that the dominant chiral phase (i.e.~$\chi_\pm$ outside and inside the CV, respectively) always winds according to the vorticity $m$, both locally around each vortex defect, and globally around the perimeter of the disc. For example, consider positive external flux and a vortex defect located at $(x,y)=(x_0,y_0)$ with winding $m$, where $m=\mp1$ for Abrikosov vortices and antivortices, respectively, while $m=-4$ for the CV considered here. Close to $(x_0,y_0)$, the dominant phase is described by $\chi_+(\vR) \approx m\phi$ with polar angle $\phi$. Far from all the vortices at the disk perimeter, the dominant chiral phase globally winds $\chi_+(\vR) \approx m_{\mathrm{tot}}\phi$, with total vorticity $m_{\mathrm{tot}}=-(N_{\mathrm{V}} + 4N_{\mathrm{CV}}) + N_{\mathrm{AV}}$, where $N_{\mathrm{V}}$ counts the number of vortices, $N_{\mathrm{CV}}$ the number of CVs, and $N_{\mathrm{AV}}$ the number of antivortices. Thus, from top to bottom row, $m_{\mathrm{tot}}=-(1+4)$, $m_{\mathrm{tot}}=-4+1$, $m_{\mathrm{tot}}=-(4+1)$, $m_{\mathrm{tot}}=-(4+2)$, $m_{\mathrm{tot}}=-(4+3)$, $m_{\mathrm{tot}}=-(4+4)$. We also find that the winding constraint $p=m+2M$ from Eq.~(\ref{eq:phase_constraint}) for the subdominant chiral phase is always fulfilled.

In Fig.~\ref{fig:cv_v:LDOS} we display the spatially-resolved LDOS for the exact same systems and solutions as in Fig.~\ref{fig:cv_v:quantities}, where each column is taken for a different fixed subgap energy $\varepsilon$ (i.e.~bias voltage). Importantly, at low energies, each Abrikosov vortex appears as a point-like peak representing the Caroli-de-Gennes-Matricon states \cite{Caroli:1964,Rainer:1996,Timmermans:2016,Berthod:2017,Kim:2021:arxiv}, which expands to a size of roughly $\sim1\xi_0$ at higher energies. By contrast, the CV appears like a ring-like peak that is an order of magnitude larger already at zero energy, $\RCV \sim 10\xi_0$. The CV expands into two concentric rings at higher energies, corresponding to the combined superflow generated by vorticity and the edge modes on either side of the domain wall. The intensity of the subgap states in the CV and Abrikosov vortex are also separated by an order of magnitude, but the LDOS peak of the CV should still be observable as it can be significantly larger than the coherence peak and is tunable by both temperature and flux, as shown in our earlier work \cite{Holmvall:2023:arxiv}.
Notably, as the CV is deformed by the Abrikosov vortices, we also see how the LDOS is correspondingly deformed in rows 2-6. Thus the LDOS is explicitly tracking the CV shape.

Figure~\ref{fig:cv_v:ASB:LDOS} shows the LDOS for similar combinations of a CV with Abrikosov vortices, but now for a parallel CV ($\Phiext<0$ such that $m=+4$ and $p=8$ instead of $m=-4$ and $p=0$), and without the antivortex scenario. For completeness, Appendix~\ref{app:cv_v_interaction} contains a plot of the corresponding order parameter amplitudes and phases for these scenarios (i.e.~the analogue of Fig.~\ref{fig:cv_v:quantities}). The overall trend in Fig.~\ref{fig:cv_v:ASB:LDOS} is similar to that of the antiparallel CV, but importantly, we note that the distinct LDOS signature of the axial symmetry breaking is clearly present, including the eightfold symmetry related to $p=m+2M=8$ for the CV (instead of $p=0$ for the antiparallel CV). Hence, despite the strong local perturbation caused by the presence of Abrikosov vortices, the overall Doppler shift caused by finite superflow from the $p=8$ winding centers remains clearly distinguishable, as is thus then the direct signature of the Chern number $M$. Interestingly, the last row of Fig.~\ref{fig:cv_v:ASB:LDOS} was initialized with four vortices outside the CV (i.e.~the same arrangement as the last row Fig.~\ref{fig:cv_v:LDOS}), but during the self-consistency loop, one of the vortices was spontaneously absorbed into the center of the CV, thereby lowering the free energy. Notably, this is a self-consistent and robust solution, illustrating that it is realistic to study and expect the appearance of configurations with an Abrikosov vortex trapped inside the CV. Importantly, during the trapping of the Abrikosov vortex in the self-consistency loop, the vortex was at some point located at the domain wall of the CV where it could have been disassociated into fractional vortices, thus leading to a higher quantized CV with $m=+5$ and $p=+9$. However, the displayed solution with $m=4$ and $p=8$ was still preferred. Hence, this is another strong indication that the quadruply quantized CV is the most robust CV in chiral $d$-wave superconductors.

Finally, on more general grounds, we point out that studying a combination of CVs and Abrikosov vortices is relevant, since both are robust topological defects and can thus appear simultaneously in a sample. This is further supported by the high energy barriers associated with vortex dynamics, meaning that a particular vortex solution can be trapped in the system far into its metastable regime \cite{Bean:1964,Burlachkov:1991,Volovik:2015:a,Benfenati:2020:b,Maiani:2022, SuperConga:2023}, where another vortex solution technically has a lower energy but can still not enter the system.
Generally, both Abrikosov vortices and domain walls can be ``kicked'' into the system by e.g.~annealing and rapid quenches in temperature and flux, and they can be further stabilized and trapped by pinning centers and certain geometry \cite{Garaud:2014,Vadimov:2013}. Indeed, we find that combinations of CVs and Abrikosov vortices spontaneously enter and stabilize in our self-consistency calculations for different flux-temperature combiniations.

In summary, these results show a robustness of the CV in the presence of Abrikosov vortices. At the same time, a tunability of the shape is demonstrated, although the CV shape can still be traced back to its original circular (octagonal) shape for the antiparallel (parallel) CV. Specifically, the LDOS at different energies appear as concentric and convex (concave) line segments, corresponding to the Doppler shifts caused by the axisymmetric (non-axisymmetric) superflow, which in turn is generated by the internal (external and internal) phase windings for the antiparallel (parallel) CV. We also note that these results give rise to an even stronger experimental signature in the LDOS, as the point-like Abrikosov vortex is distinctly different from the line-like CV. Finally, we propose that similar deformations might be caused by other strong local electromagnetic perturbations, e.g.~an appropriately prepared STM tip with strong magnetization.

\section{Non-circular geometry and strong confinement}
\label{sec:non_circular_grains}
The previous two sections \ref{sec:cv_size} and \ref{sec:cv_v_interaction} illustrated that the overall size and shape of the CV is balanced by effective attractive versus repulsive electrodynamical interactions, traced back to the domain wall currents and fractional vortices respectively. In this section, we further illustrate this via the interaction with the system edges, and show how confinement alone can induce asymmetric deformation modes in the CV. In addition, the results show that the LDOS signature remains robust and is not relying on the symmetry (or lack thereof) of the system itself.

\begin{figure*}[bt!]
	\includegraphics[width=\textwidth]{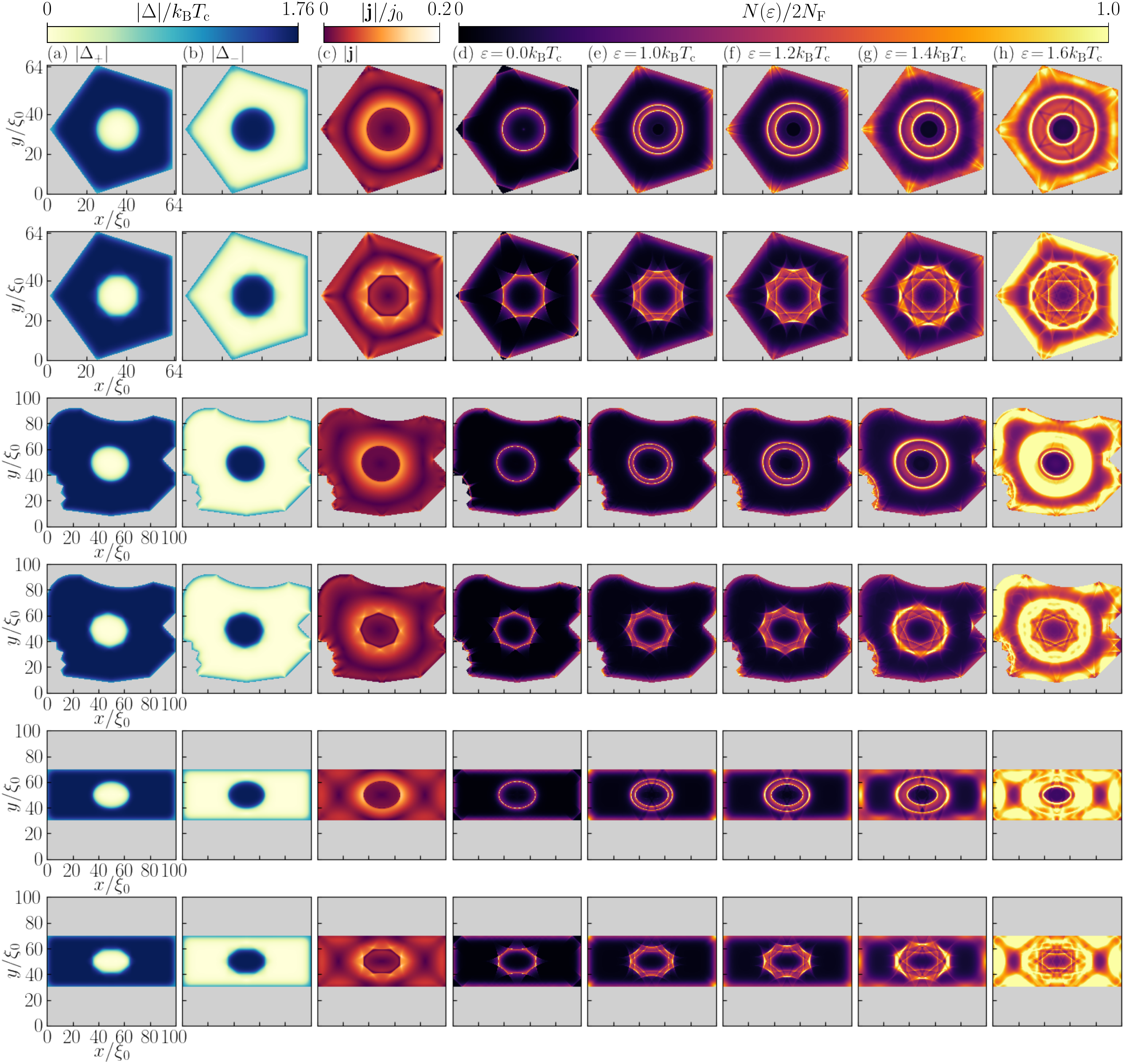}
	\caption{CVs in various non-circular samples, with antiparallel (parallel) CVs in odd (even) rows. Columns, from left to right, show magnitude of the chiral order parameter components, charge-current density, and  LDOS at different fixed subgap energies. Here, dominant bulk chirality is $\Dplus$, with $T=0.1\Tc$, $\lambda_0=80\xi_0$, and with $\Phiext=\pm8\Phi_0$ for antiparallel and parallel CVs respectively. }
	\label{fig:non_circular_grain}
\end{figure*}

Figure~\ref{fig:non_circular_grain} shows an antiparallel (parallel) CV in odd (even) rows in systems with different shapes, where the columns show from left to right: magnitude of the chiral order parameter components, charge-current density, and LDOS at different fixed subgap energies. The first two rows show a sample shaped like a pentagon, importantly illustrating that the overall circular versus octagon rotation symmetries of the CVs remain, even when incommensurate with the rotation symmetry of the system. Furthermore, this is an example of a system with higher-order harmonics discussed in Sec.~\ref{sec:methods:chiral}, where Eq.~(\ref{eq:phase_constraint}) is modified with an additional term, such that $p=m+2M+n$, here with integer $n=-5$ due to the five-fold rotational symmetry of the superconducting grain. This leads to additional phase gradients and therefore superflow, which in turn generates additional current components. This effect is responsible for helping the current turn the sharp corners of the system, which is a well-known effect in chiral superfluids \cite{Sauls:2011}. Furthermore, the additional phase gradients and superflow also leads to a locally enhanced LDOS at the corners at finite energies, again via the Doppler shift discussed in relation to Eq.~(\ref{eq:superflow}). As a result, the subdominant phase $\chi_-$ of the antiparallel CV has integer winding $p=-4+4+n=-5$, while $p=4+4+n=3$ for the parallel CV. Hence, the higher-order harmonics is superimposed with the antiparallel versus parallel vorticity and chirality, especially seen by the additional signatures with five-fold rotational symmetry in the LDOS at high energies in the last column Fig.~\ref{fig:non_circular_grain}(h). Importantly, the higher-order harmonics still does not modify the overall strong signature of vorticity and chirality in the LDOS. In other words, the strong LDOS distinction between parallel and antiparallel CVs remains robust.
Next, the third and forth rows show a completely irregular system without any rotation symmetry. Again, the LDOS signature is robust, but the sharp wedges together with the overall asymmetry between $x$ and $y$ directions cause a slight deformation of the CVs. The last two rows show a rectangular system, with an even stronger asymmetry between $x$ and $y$ directions. Due to the effectively repulsive interaction between the system edges and the fractional vortices in the CV, the resulting CV shape is strongly deformed, with a clear $x$ and $y$ asymmetry. The effective repulsive interaction with the system edges is related to the energy barriers for vortex entrance and expulsion at sufficiently high external flux \cite{Bean:1964,Burlachkov:1991,Volovik:2015:a,Benfenati:2020:b,Maiani:2022, SuperConga:2023}.
In summary, this section illustrates both a tunability of the CV shape due to mesoscopic confinement, and most importantly that despite these CV shape changes, the experimental signatures in the LDOS are robust at all subgap energies and do not rely on the overall rotation symmetry of the system.

\section{Non-circular Fermi surfaces}
\label{sec:non_circular_fs}
\begin{figure}[bt!]
	\includegraphics[width=\columnwidth]{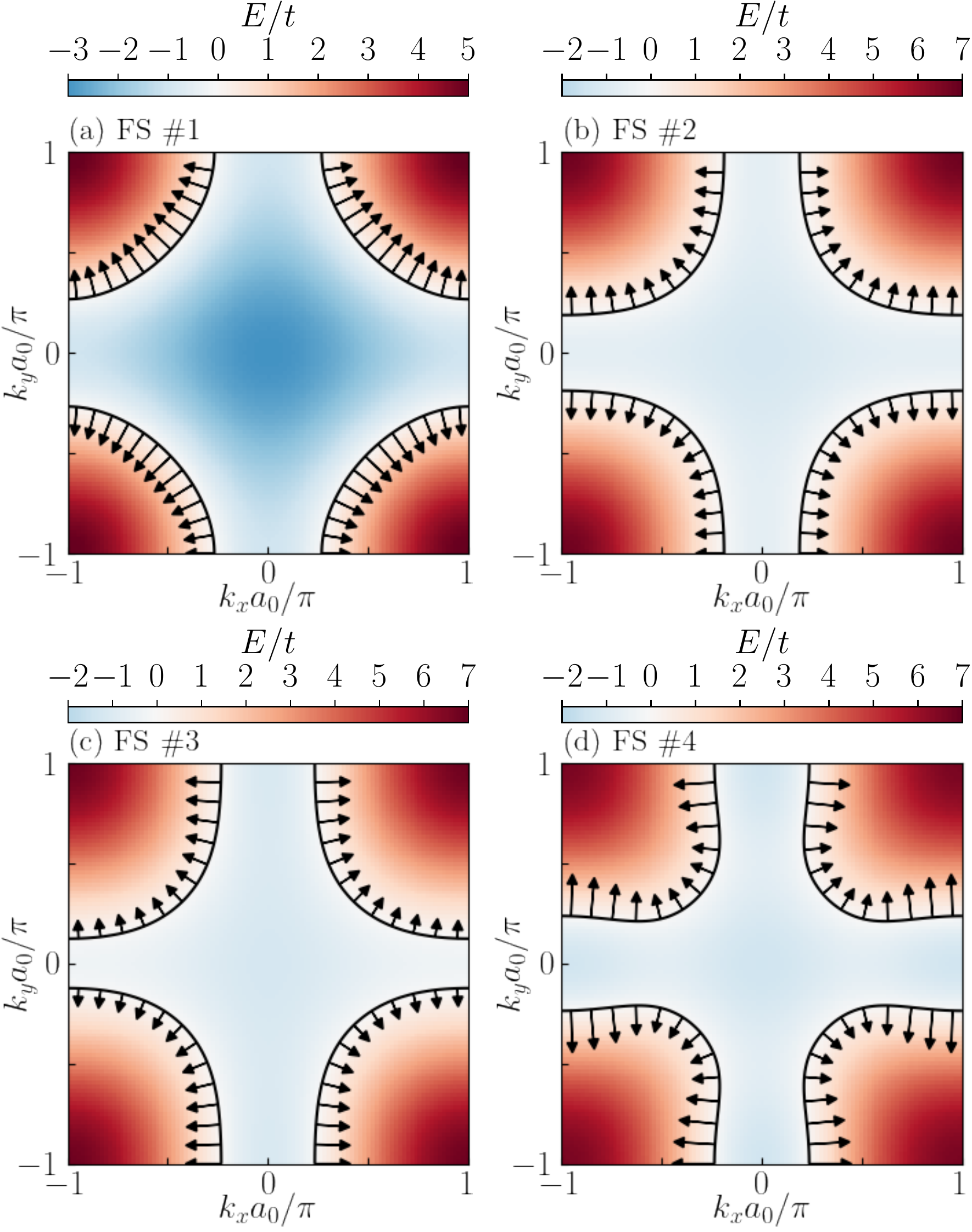}
	\caption{Normal-state band structures for the non-circular tight-binding hole-doped FSs defined in Table.~\ref{tab:fermi_surfaces}. Colors indicate band energy $E$, solid lines denote FS ($E=0$), with arrows indicating the Fermi velocity $\vvF(\vpF)$ used as input in the quasiclassical parametrization \cite{Wennerdal:2020}. }
	\label{fig:non_circular_FS}
\end{figure}

So far, we have assumed a circular and electron-doped FS as in our previous work \cite{Holmvall:2023:arxiv}. Here we show that our main results and conclusions do not depend on the shape of the FS or particular doping level. In particular, we consider FSs formed in a hole doped material and with weak to strong deviation from a circular shape, and also with anisotropy between $k_x$- and $k_y$-momentum directions, to further mimic possible broken symmetries in the normal state. In particular, we parametrize a non-circular FS via the momentum $\mathbf{k} = k_x\hat{k}_x + k_y\hat{k}_y$ through the normal-state dispersion $\epsilon_{\mathbf{k}}$ on a square lattice
\begin{align}
    \nonumber
    \epsilon_{\mathbf{k}} = &-2t[(1+\alphaxy)\cos(k_xa_0) + (1-\alphaxy)\cos(k_ya_0)] \\
    \nonumber
    &- 4t^{\prime}\cos(k_xa_0)\cos(k_ya_0)\\
    \label{eq:square_dispersion}
    &-2t^{\prime\prime}[(1+\alphaxy)\cos(2k_xa_0) + (1-\alphaxy)\cos(2k_ya_0)],
\end{align}
in terms of the lattice constant $a_0$, nearest-neighbor hopping $t>0$ (which we use as a natural unit for all tight-binding energies), next-nearest neighbor hopping $t^{\prime}$, next-next nearest neighbor hopping $t^{\prime\prime}$, and with anisotropy $\alphaxy$ between $k_x$ and $k_y$ \cite{Wahlberg:2021}. We here consider four different tight-binding models taken from the literature \cite{Radtke:1994,Meevasana:2008,Berthod:2017}, labeled as FS $\#1$ to $\#4$, defined in Table~\ref{tab:fermi_surfaces} and illustrated in Fig.~\ref{fig:non_circular_FS}.
\begin{table}[h!]
\def\arraystretch{1.3}
\begin{ruledtabular}
\begin{tabular}{c | c | c | c | c}
    FS & $t^{\prime}$ & $t^{\prime\prime}$ & $\mu$ & $\alphaxy$ \\
\hline
    $\#1$ & $-0.250t$ & $0$ & $0$ & $0$ \\
    $\#2$ & $-0.437t$ & $0.034t$ & $-1.203t$ & $0$ \\
    $\#3$ & $-0.437t$ & $0.034t$ & $-1.203t$ & $0.1$ \\
    $\#4$ & $-0.495t$ & $0.156t$ & $-1.267t$ & $0$ \\
\end{tabular}
\end{ruledtabular}
\caption{Parametrization of tight-binding Fermi surfaces (FSs) for the normal-state dispersion in Eq.~(\ref{eq:square_dispersion}) with nearest neighbor hopping $t$, next-nearest neighbor hopping $t^{\prime}$, next-next nearest neighbor hopping $t^{\prime\prime}$, chemical potential $\mu$, and hopping anisotropy $\alphaxy$. Resulting FSs are illustrated in Fig.~\ref{fig:non_circular_FS}.}
\label{tab:fermi_surfaces}
\end{table}
Here, the non-circular FS leads to a modified $\vvF(\vpF)$ entering the Eilenberger equation (\ref{eq:model:eilenberger}), thus modifying the propagators and all other quantities defined in Sec.~\ref{sec:methods} correspondingly. See Ref.~\cite{Wennerdal:2020} for further details on parametrizing such a microscopic tight-binding Fermi surface within quasiclassical theory of superconductivity. We further note that all of these FSs are hole-doped, corresponding to being centered around $(k_x,k_y) = (\pi/a_0,\pi/a_0)$, and show either a four-fold (FSs $\#1$, $\#2$, $\#4$) or two-fold (FS $\#3$) discrete rotational symmetry. There is a weak to strong deviation from circular shape in changing between FS $\#1$ to FS $\#4$. In contrast, an electron-doped FS is centered around $(k_x,k_y) = (0,0)$.

\begin{figure*}[bt!]
	\includegraphics[width=\textwidth]{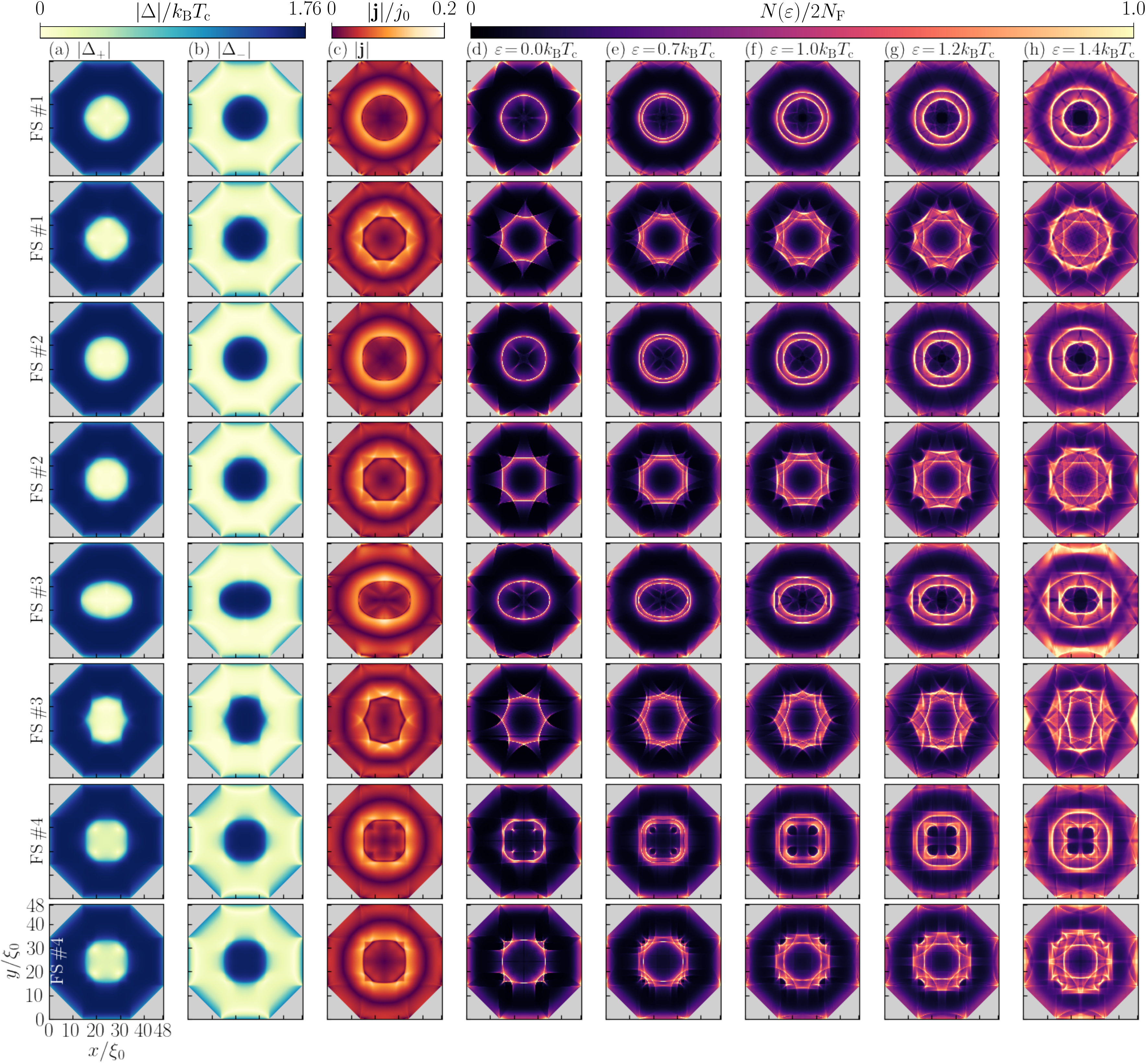}
	\caption{CVs in an octagon-shaped sample. Different rows show the different FSs defined in Table~\ref{tab:fermi_surfaces}, FS $\#1$ to FS $\#4$, with antiparallel (parallel) CV in odd (even) rows. Columns, from left to right, show magnitude of the chiral order parameter components, charge-current density, and LDOS at different fixed subgap energies. Here, $\Dplus$ is the dominant bulk chirality with $T=0.1\Tc$, $\lambda_0=80\xi_0$, and with $\Phiext=\pm8\Phi_0$ for antiparallel and parallel CVs respectively.}
	\label{fig:non_circular_FS:octagon}
\end{figure*}
\begin{figure*}[bt!]
	\includegraphics[width=\textwidth]{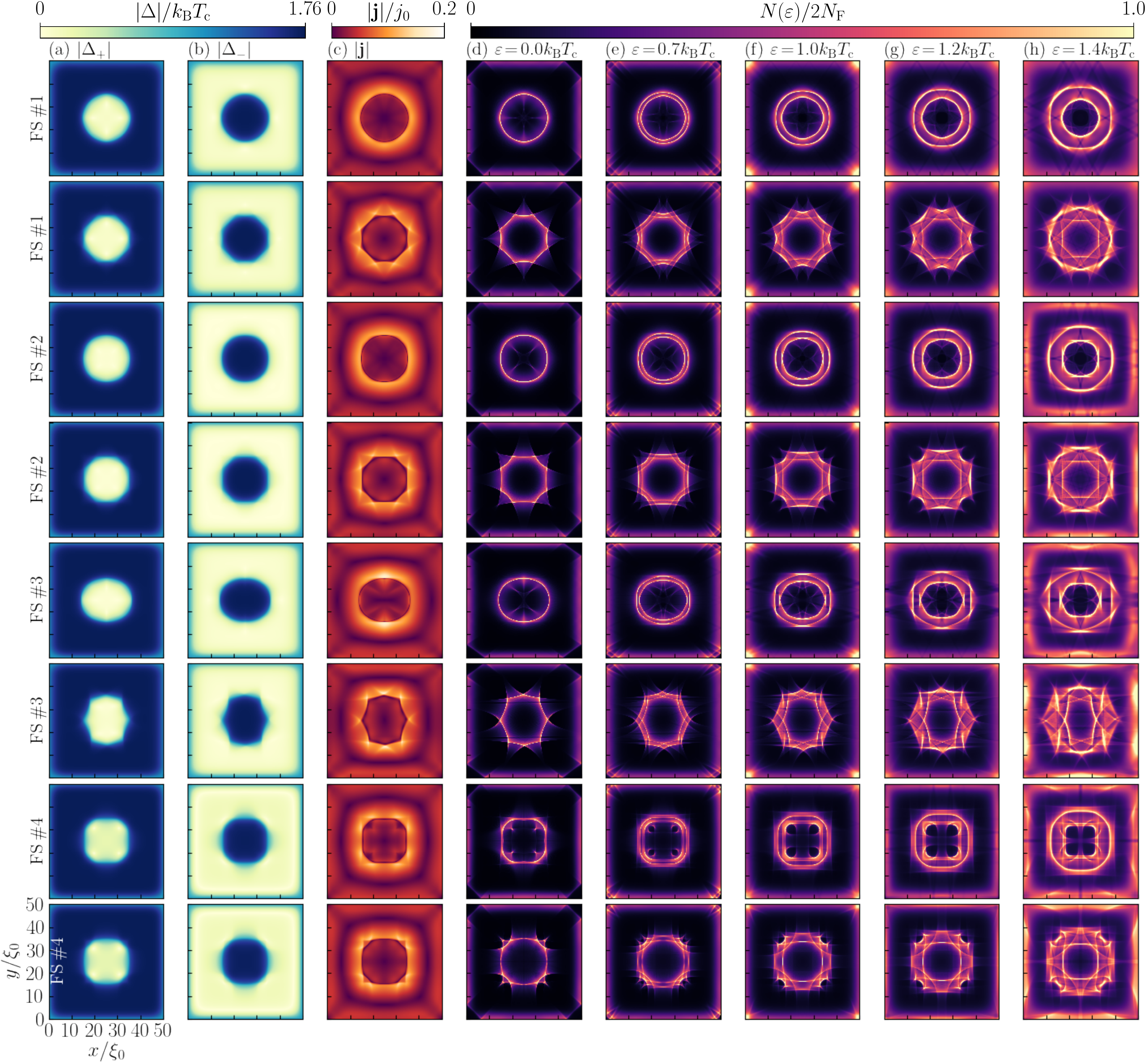}
	\caption{Same as Fig.~\ref{fig:non_circular_FS:octagon} but for a square-shaped sample.}
	\label{fig:non_circular_FS:square}
\end{figure*}

We show the antiparallel (parallel) CV computed with these FSs in odd (even) rows in Fig.~\ref{fig:non_circular_FS:octagon} for an octagonal sample and in Fig.~\ref{fig:non_circular_FS:square} for a square sample. In all figures, the columns show from left to right the chiral order parameter amplitudes, charge-current density, and LDOS at different subgap energies $\varepsilon$.
Overall, we find that both types of CVs show traces of the underlying symmetry of the FS, which can be explained in terms of higher-order harmonics superimposed on the CV as discussed in Sec.~\ref{sec:methods:chiral} and studied for non-circular samples in Sec.~\ref{sec:non_circular_grains}, but here they are instead originating from the FS. For example, a four-fold rotational symmetry of the FS (or sample) leads to a corresponding four-fold rotational symmetry with nodes and kinks developing in the CV. Similarly, an anisotropic FS with two-fold rotational symmetry (FS $\#3$) deforms the otherwise circular CV into an ellipse, due to suppression (enhancement) of $\vvF$ along $k_y$ ($k_x$), as illustrated in Fig.~\ref{fig:non_circular_FS}(c). Interestingly, the elliptical deformation occurs along opposite directions for the antiparallel and parallel CVs, which we interpret to be due to opposite signs of $\vvF(\vpF)\cdot\vps(\vR)$ for the two CVs, which can be traced back to opposite signs of $\vA$ (i.e.~opposite external field directions) entering $\vps$ in Eq.~(\ref{eq:superflow}). Furthermore, we note here that FS $\#4$ is very distorted compared to a circular FS, leading to also very strong distortions in the CV.

Despite these symmetry-breaking terms in the FS causing distortion of the CV, we find that both CV solutions are always robust, and the important asymmetry remains clear in the LDOS. Specifically, the CV with antiparallel vorticity and chirality (odd rows) generates convex and concentric lines in the LDOS, from the axisymmetric angular momentum and superflow. In contrast, the CV with parallel vorticity and chirality (even rows) always generates characteristic concave LDOS patterns due to the multiple phase winding centers which are non-overlapping (i.e.~axial symmetry-breaking).
Moreover, the emergent discrete rotational symmetry and interweaving resonances at higher energies is a direct experimental signature of the quantized vorticity and Chern number, due to it tracing back to winding superposition $p=m+2M$, which is robust due to quantization and topology \cite{Holmvall:2023:arxiv}. These results establish that the signatures of CVs are robust even for a highly anisotropic FS, reflecting broken symmetries of the normal state.

\section{Non-degenerate nodal components}
\label{sec:non_degeneracy}

\begin{figure*}[tb!]
	\includegraphics[width=\textwidth]{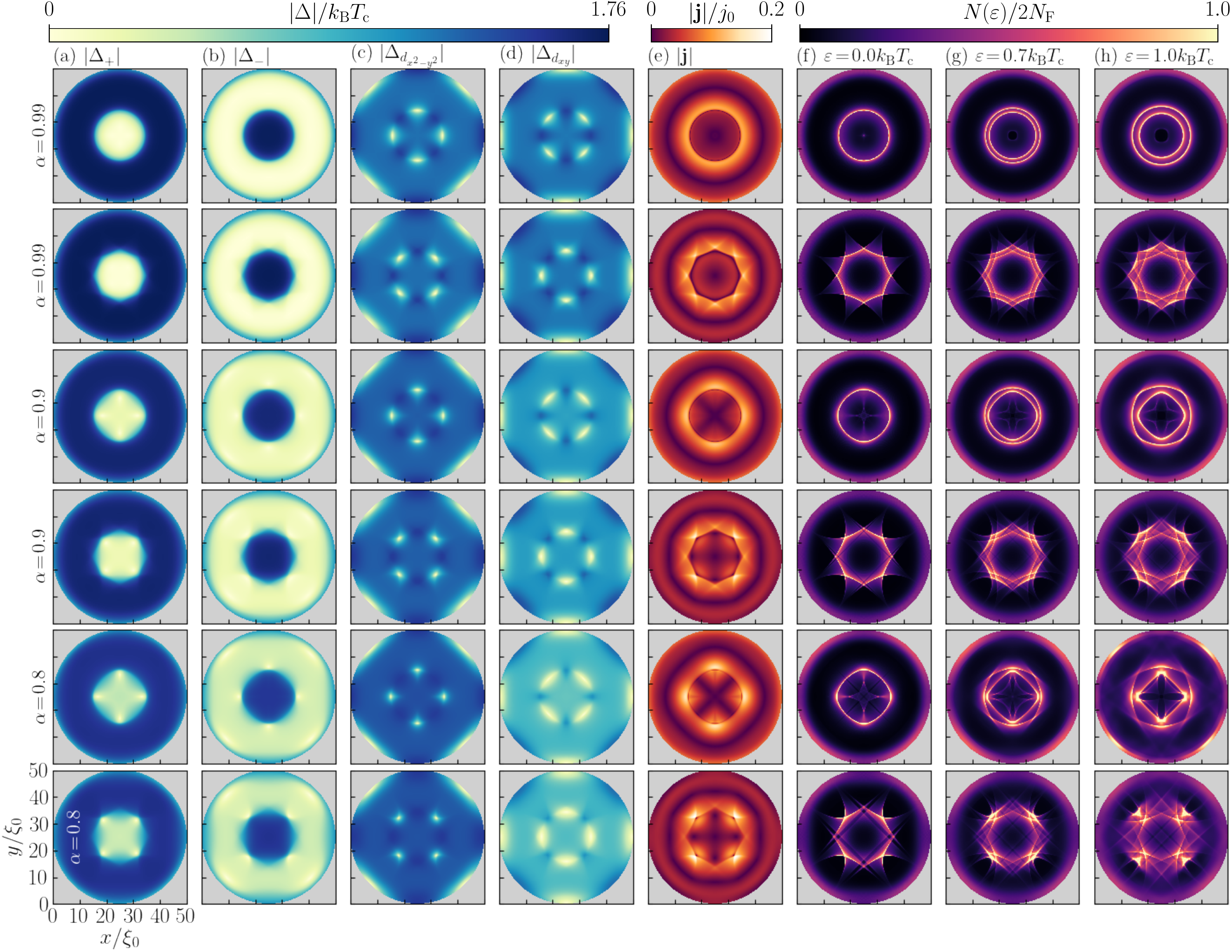}
	\caption{CVs with broken degeneracy between the two nodal order parameter components, quantified by the $\Tc$ ratio $\alpha$ in Eq.~(\ref{eq:alpha_Tc}), in a system with dominant bulk chirality $\Dplus$ at $T=0.1\Tc$, $\lambda_0=80\xi_0$, and $\mathcal{R}=25\xi_0$. Antiparallel (parallel) CVs in odd (even) rows corresponding to $\Phiext=+8\Phi_0$ ($\Phiext=-8\Phi_0$).}
	\label{fig:non_degenerate_CV}
\end{figure*}

In all other sections and our previous work \cite{Holmvall:2023:arxiv} we assumed degeneracy between the two nodal $d$-wave pairing symmetries, $d_{x^2-y^2}$ and $d_{xy}$, such that their transition temperatures are the same. Such an exact degeneracy is experimentally relevant: it is enforced by the symmetry and group theory in any material with a three- or six-fold rotationally symmetric lattice \cite{Black-Schaffer:2014:b}. This includes triangular, hexagonal, and honeycomb materials and is as such guaranteed in many of the materials currently proposed as chiral $d$-wave superconductors \cite{Venderbos:2018,Su:2018,Fidrysiak:2018,Xu:2018,Kennes:2018,Liu:2018,Gui:2018,Wu:2019,Fischer:2021,Ming:2023,Biswas:2013,Fischer:2014,Ueki:2019,Ueki:2020}.
Still, for sake of full completeness, we in this section show that our main results and conclusions in addition hold for systems where this degeneracy is somehow broken. Specifically, we consider non-degenerate pairing interactions modeled by different coupling constants resulting in different transition temperatures, quantified by the ratio
\begin{equation}
    \label{eq:alpha_Tc}
    \alpha \equiv \frac{\Tcdxy}{\Tcdxtyt} \in [0,1].
\end{equation}
Hence, we set the $d_{xy}$-component to be subdominant for all $\alpha < 1$, resulting also in different bulk amplitudes $|\Dxy| < |\Dxtyt|$. However, apart from inserting these different coupling strengths, we do not constrain the order parameter components in any way, and solve for both of them completely self-consistently. For example, performing self-consistent calculations without any vorticity, we still find that the chiral $d$-wave state is the ground state even for highly non-degenerate systems with $\alpha < 0.8$, thus surviving a strong suppression of the $d_{xy}$-component. Notably, such a state is still fully gapped in the bulk, with a Chern number $M = \pm 2$. Hence, the possibility of antiparallel versus parallel superposition of vorticity and chirality in a CV is still possible.

In order to investigate the effects on the CV from non-degenerate $d$-wave nodal components, we begin by summarizing the scenario of full degeneracy ($\alpha=1$) studied so far. Here, both the axially symmetric CV with antiparallel vorticity and chiralty, and the axial symmetry-breaking CV with parallel vorticity and chirality, are extremely robust solutions over a large range of temperatures and flux. Notably, for degenerate nodal $d$-wave components in a disc-shaped system and FS, the total superconducting order parameter has full rotation symmetry for the antiparallel CV, and thus physical properties such as currents and magnetic fields generally do not reflect the four-fold symmetry of the individual nodal components. 
However, as the degeneracy between the nodal components is broken, it is reasonable to expect that the nodal four-fold rotational symmetry will be imprinted also on the antiparallel CV.

In Fig.~\ref{fig:non_degenerate_CV} we study antiparallel and parallel CVs from weak non-degeneracy $\alpha=0.99$ (top two rows) and continue to strong non-degeneracy $\alpha=0.8$ (lowest two rows). For this full range of asymmetry, we find that both CVs are still very robust, but over a slightly narrower range of flux. By decreasing $\alpha$ we find that the broken degeneracy and four-fold nodal symmetry become more apparent in the CV, as expected. For example, along the domain wall, the suppression of $\Dxtyt$ ($\Dxy$) now occurs over a smaller (larger) region, as compared to $\alpha=1$. As $\alpha$ is further reduced, the dominant $\Dxtyt$ covers nearly the whole domain wall, except at four isolated points. Consequently, these points become the only locations in the domain wall where $\Dxy$ is finite. The fractional vortices are however still well-separated, and the CV structure is notably still intact. This spatial structure of the individual nodal components leads to signatures also in all other quantities, including the chiral order parameters components, and also the currents, induced flux (not shown here), and LDOS. Still, the results in Fig.~\ref{fig:non_degenerate_CV} show that the overall conclusions and experimental signatures established in the rest of the work for the degenerate case remain robust and clear also with a strong asymmetry between the two nodal $d$-wave components.  In particular, the LDOS for the antiparallel CV keeps its overall concentric and convex circular lines due to the order parameters and currents also exhibiting such a profile, with non-degeneracy only turning the circles more square-like. Meanwhile, the parallel CV still shows concave octagonal structures, with eight-fold interweaving resonances at higher energy due to non-trival additional phase winding in the subdominant chirality (but now overlapped with strong four-fold structure). Thus the antiparallel CV keeps the axisymmetry, while the parallel CV does not, just as established in the rest of the results.
This robustness in the different LDOS patterns between the two CVs is expected: after all, the LDOS patterns stem directly from a positive versus negative superposition of the quantized and topologically protected Chern number (OAM from chirality) and vorticity (c.m.~angular momentum), as introduced in Sec.~\ref{sec:methods:chiral}. Thus, as long as there is a chiral state, the positive versus negative superposition generates completely different scenarios, but possibly superimposed with higher-order contributions, in this case due to the additional broken nodal degeneracy.

Finally, for completeness, let us address the extreme limit of non-degeneracy, although this is not expected for stable chiral $d$-wave superconductors and thus not of central importance here. Eventually, the local variation of the current leads to a modified line tension, modifying the stability. Indeed, as $\alpha \to 0.6$, the parameter-space region of stable CV shrinks rapidly. At some point, the CV also becomes unstable and multiple regular Abrikosov vortices are instead stabilized. We define the critical ratio where this occurs as $\alpha^*(T,\Phiext,\lambda_0,\mathcal{R})$, hence possibly depending on all parameters such as temperature, flux, penetration depth, and system size. The full parameter space is of course far beyond the scope of the present work, but we consider a subset of the parameter space for illustrative purposes. For example, at fixed temperature $T=0.1\Tc$, external flux $|\Phiext|=8\Phi_0$, and $\lambda_0=80\xi_0$, we find that the antiparallel CV is unstable below $\alpha^*\approx0.70$ at $\mathcal{R}=25\xi_0$, $\alpha^*\approx0.60$ at $\mathcal{R}=50\xi_0$, and $\alpha^*\approx0.55$ at $\mathcal{R}=75\xi_0$, while the parallel CV is unstable below $\alpha^*\approx0.73$ at $\mathcal{R}=25\xi_0$, $\alpha^*\approx0.64$ at $\mathcal{R}=50\xi_0$, and $\alpha^*\approx0.61$ at $\mathcal{R}=75\xi_0$. The CV is thus less stable in smaller systems, especially for increased non-degeneracy. We interpret this to stem from that smaller systems exhibit significant overlap between opposite system edges where both of the nodal components are suppressed, as well as between the CV and the system edge. This suppression is naturally enhanced by the non-degeneracy. As a result, the chiral state competes with both the normal state and a nodal $d$-wave state, which effectively hampers the formation of the chiral state, and consequently therefore also the formation of domain walls and CVs.

\section{Non-magnetic impurities}
\label{sec:impurities}
Our earlier work \cite{Holmvall:2023:arxiv} demonstrated that the LDOS signatures of the CV are robust under the inclusion of a phenomenological energy broadening $\delta$ of the spectrum. Such an energy broadening can be caused by e.g.~disorder, impurity scattering, fluctuations, or interfaces with finite transparency \cite{Poenicke:1999,Lofwander:2001,Seja:2019:lic,Seja:2021,Seja:2022:thermopower,Seja:2022:current_injection}. Here we exemplify this by studying dirty systems with non-magnetic impurities, and show that the CV as well as the LDOS signatures, are robust. 

\begin{figure}[tb!]
	\includegraphics[width=\columnwidth]{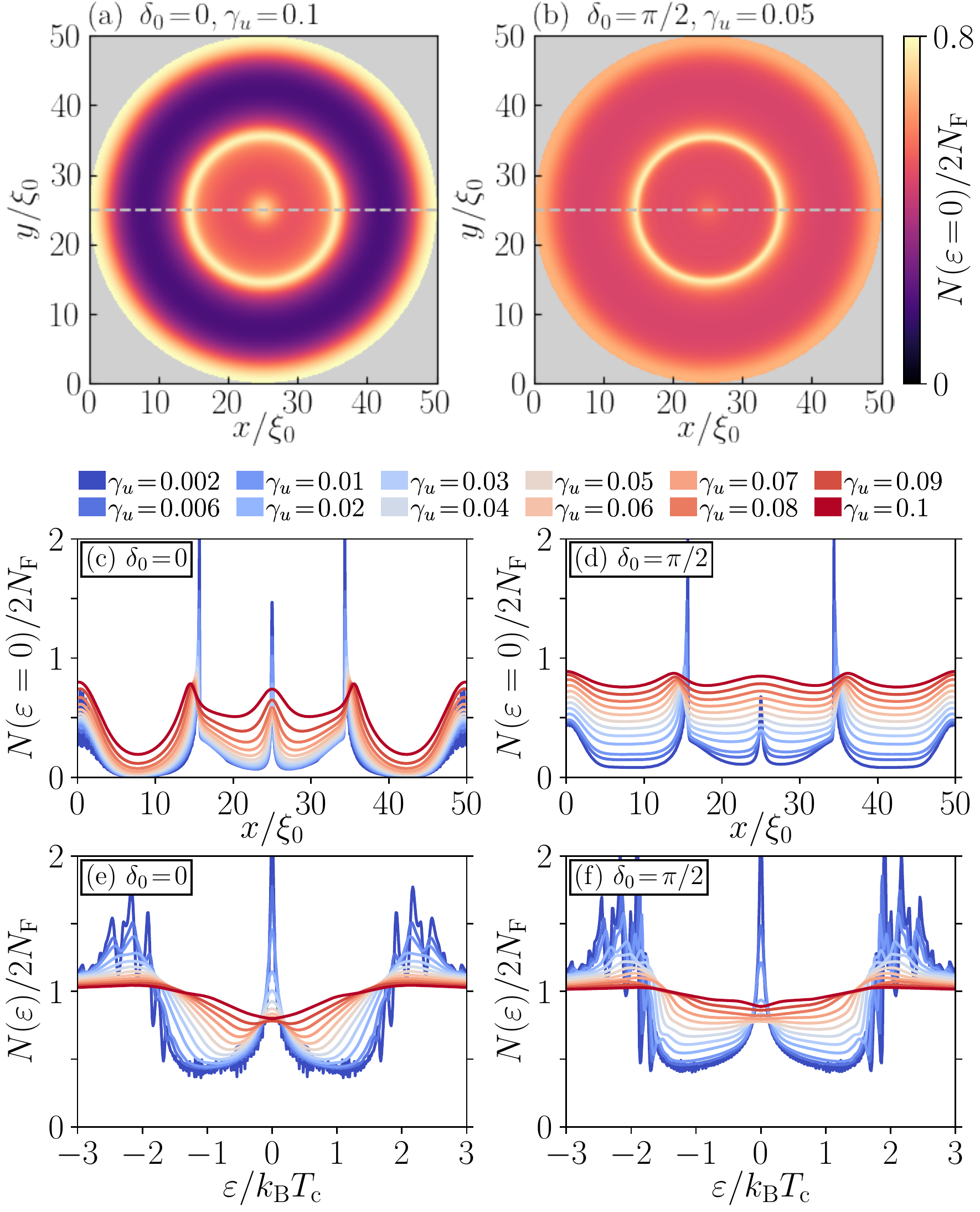}
	\caption{LDOS for an antiparallel CV in a system with non-magnetic impurities, with dominant bulk chirality $\Dplus$, $T=0.1\Tc$, $\Phiext=8\Phi_0$, and $\lambda_0=80\xi_0$. Left (right) column corresponds to the Born limit (unitary limit) with scattering phase shift $\delta_0=0$ ($\delta_0=\pi/2$). (a,b) Zero-energy LDOS. (c,d) Zero-energy LDOS across horizontal dashed line in (a,b). (e,f) LDOS in the domain wall. Line colors in (c-f) denote $\gamma_u\equiv\Gamma/(2\pi\kB\Tc)$.}
	\label{fig:impurity_LDOS:antiparallel}
\end{figure}
\begin{figure}[tb!]
	\includegraphics[width=\columnwidth]{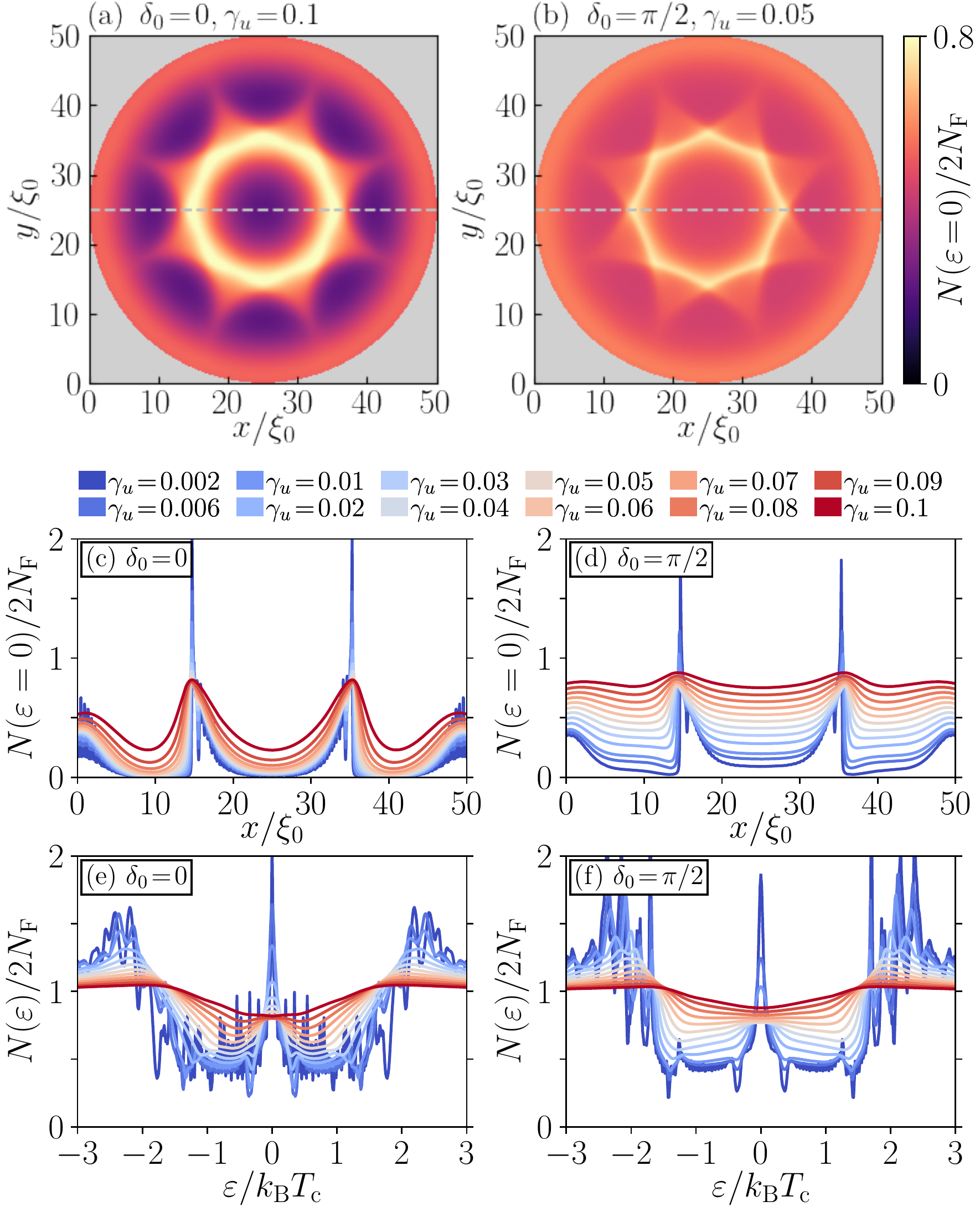}
	\caption{Same as Fig.~\ref{fig:impurity_LDOS:antiparallel}, but for a parallel CV with $\Phiext=-8\Phi_0$.}
	\label{fig:impurity_LDOS:parallel}
\end{figure}

We model the non-magnetic impurities using the well-established $t$-matrix approach within the quasiclassical theory of superconductivity \cite{Graf:1996,Seja:2022:thesis}. The diagonal impurity self-energy from Eq.~(\ref{eq:model:self_energy:diagonal}) then takes the form
\begin{align}
    \label{eq:self_energy:impurity}
    \hat{\Sigma}_{\mathrm{imp}}(\vpF,\vR;z) = & n_i\hat{t}(\vpF,\vpFp\to\vpF,\vR;z),
\end{align}
with dilute impurity concentration $n_i$ and impurity-scattering matrix $\hat{t}$ fulfilling the additional self-consistency equation
\begin{align}
    \nonumber
    \hat{t}(\vpF,\vpFp,\vR;z) = & \NF\Big\langle \hat{u}_0(\vpF,\vpFpp;z) \hat{g}(\vpFpp,\vR;z)\\
    \label{eq:self_energy:t}
    & \times\hat{t}(\vpFpp,\vpFp,\vR;z)\Big\rangle_{\vpFpp}
    + \hat{u}_0(\vpF,\vpFp;z),
\end{align}
with scattering potential $\hat{u}_0(\vpF,\vpFp;z)$. Equation~(\ref{eq:self_energy:t}) results from a diagrammatic expansion describing multiple scattering of quasiparticles and pairs by an impurity, connecting different scattering channels with momenta $\vpF$ and $\vpFp$ on the FS (here integrated over the momentum $\vpFpp$). Here we assume equilibrium, a non-crossing approximation, and an isotropic scattering potential, such that $\hat{u}_0(\vpF,\vpFp;z) = u_0\hat{1}$, yielding
\begin{align}
    \label{eq:self_energy:t:solution}
    \hat{\Sigma}_{\mathrm{imp}}(\vR;z) & = \Gamma_u\frac{\sin\delta_0\cos\delta_0\hat{1} + \sin^2\delta_0\left\langle\frac{1}{\pi} \hat{g}(\vpF,\vR;z)\right\rangle_{\vpF}}{\cos^2\delta_0\hat{1} - \sin^2\delta_0\left(\frac{1}{\pi}\left\langle \hat{g}(\vpF,\vR;z)\right\rangle_{\vpF}\right)^2},
\end{align}
with scattering energy $\Gamma_u = n_i/(\pi\NF)$ and scattering phase shift $\delta_0 = \arctan(\pi u_0 \NF)$. We solve Eq.~\eqref{eq:self_energy:t:solution} self-consistently, together with the gap equation and Maxwell's equation. We define the ``pair-breaking energy'' as $\Gamma=\Gamma_u\sin^2\delta_0$, related to the normal-state mean-free-path $l=\hbar\vF/(2\Gamma)$. We consider two extreme limits, namely the weak-scattering Born limit ($\delta_0 \to 0$ and $\Gamma_u \to \infty$, such that $\Gamma$ is constant) and the strong-scattering unitary limit ($\delta_0\to\pi/2$ and $u_0 \to \infty$, such that $\Gamma = \Gamma_u$). In these limits, the equilibrium solutions simplify to 
\begin{align}
    \label{:self_energy:t:solution:Born}
    \hat{\Sigma}_{\mathrm{imp}}^{\mathrm{Born}}(\vR;z) & = \frac{\Gamma}{\pi}\langle \hat{g}(\vpF,\vR;z) \rangle_{\vpF},\\
    \label{:self_energy:t:solution:unitary}
    \hat{\Sigma}_{\mathrm{imp}}^{\mathrm{unitary}}(\vR;z) & = -\pi\Gamma\frac{\langle \hat{g}(\vpF,\vR;z) \rangle_{\vpF}}{\langle \hat{g}^2(\vpF,\vR;z) \rangle_{\vpF}},
\end{align}
respectively. We vary the scattering energy over orders of magnitude, $\gamma_u \equiv \Gamma/(2\pi\kB\Tc) \in [0.002, 0.1]$. By comparison, the zero-temperature bulk gap is roughly $|\Delta_0|/(2\pi\kB\Tc) \approx 0.280$. We still use a phenomenological broadening $\delta/(2\pi\kB\Tc) = 0.0005$ to avoid divergent LDOS for small $\gamma_u$, but this value is an order of magnitude smaller than used in the rest of this work, $\delta/(2\pi\kB\Tc) = 0.005$. 

Figures~\ref{fig:impurity_LDOS:antiparallel} and \ref{fig:impurity_LDOS:parallel} show the resulting LDOS in the presence of non-magnetic impurities for an antiparallel and parallel CV, respectively, with left and right columns showing the Born and unitary limits, respectively, with the figure panels (c-d) showing line-cuts across the CV at zero energy and panels (e-f) showing the LDOS at the domain wall.
As expected, the LDOS peaks are broadened when increasing the scattering energy, eventually becoming almost completely broadened for $\gamma_u \to 0.1$ as indicated by red lines in (c-f). This result is expected because such strong $\gamma_u$ is comparable with the bulk gap. The broadening is also naturally not as strong in the Born limit (left) as in the unitary limit (right). Consequently, panels (a,c) illustrate that both CVs are strongly distinguishable in the spatially resolved LDOS even for $\gamma_u=0.1$ in the Born limit, while panels (b,d) show that both CVs are distinguishable at $\gamma_u=0.05$ in the unitary limit. We note that the peak at the disc center in Fig.~\ref{fig:impurity_LDOS:antiparallel} is a resonance related to the perfect rotation symmetry \cite{Holmvall:2023:arxiv}.

Finally, we note that the antiparallel CV radius $\RCV$ slightly increases by $1\xi_0$ ($2\xi_0$) as $\gamma_u$ changes from $0.002$ to $0.1$ for the Born (unitary) limit, as indicated in Figs.~\ref{fig:impurity_LDOS:antiparallel}(c,d). For the parallel CV, the increase in $\RCV$ is even smaller, about $0.5\xi_0$, see Figs.~\ref{fig:impurity_LDOS:parallel}(c,d). We expect $\RCV$ to increase more significantly with $\gamma_u$ in systems where $\mathcal{R} \geq \lambda_0$. This is because non-magnetic impurities generally increase the penetration depth $\lambda_0$ \cite{Prozorov:2022}, which in turn increases $\RCV$ as shown in Sec.~\ref{sec:cv_size}. Here, in contrast, $\lambda_0=80\xi_0$ is much larger than $\mathcal{R}$, explaining the small variation in $\RCV$.
In summary, we find that the LDOS signatures of the Chern number, superconducting pairing symmetry, and chirality is robust against strong (moderate) scattering energy in the Born (unitary) limit, despite a corresponding broadening of the LDOS peaks. Furthermore, we find that the CV itself is very robust in all cases considered. This demonstrates the viability of CVs and its signatures to identify chiral superconductivity also in dirty systems.

\section{Conclusions}
\label{sec:conclusions}
In this work, we show a strong tunability of CVs in spin-singlet chiral $d$-wave superconductors, as well as a robustness of their experimental signature for a large range of material models, parameter regimes, perturbations, anisotropy, and disorder.

In terms of tunability, we find that the finite size of the CV is balanced by the attractive and repulsive interactions exerted by its domain-wall currents and fractional vortices, respectively. Thus, we show that the overall size is easily tuned directly by changing an externally applied magnetic flux and the temperature, but also depend on system size and penetration depth, the latter generally tunable by artificially adding impurities. We also find that the overall shape is tunable and deforms in an anisotropic environment, e.g.~due to other vortices, an irregular system shape, or an anisotropic FS.

For the experimental signatures, our earlier work established that the LDOS host distinct signatures for the two inequivalent CVs, with antiparallel or parallel chirality and vorticity, and that this can be used to clearly identify chirality, Chern number, and even the superconducting pairing symmetry \cite{Holmvall:2023:arxiv}. More specifically, the antiparallel CV is axisymmetric with a continuous rotation symmetry, associated with LDOS peaks appearing as concentric and convex circular lines. The parallel CV spontaneously breaks axial symmetry, generating additional winding centers outside the CV, deforming its shape into a concave structure with discrete rotation symmetry, directly related to the Chern winding number. At zero energy (bias voltage), the LDOS directly probes the domain wall structure of the CV and its overall rotation symmetry and thereby the Chern number. At higher energies, there are additional interweaving resonances between the additional winding centers, even more clearly exhibiting the rotation symmetry and Chern number. This forms strong experimental signatures, directly measurable with STS and STM. In this work we establish that all of these signatures are robust for a large range of possible perturbations, system, and model changes. In particular, we demonstrate how the results hold in systems with incommensurate or no rotation symmetry, at strong confinement, for both electron doped and hole doped FSs or anisotropic FSs, non-degenerate nodal $d$-wave components, as well as when non-magnetic impurities are present. We also find robustness for large ranges of temperatures, external flux strength, penetration depths, and system sizes, as well as when additional Abrikosov vortices are present. 

In conclusion, our work establishes CVs as a tunable and robust experimental signature of spin-singlet chiral $d$-wave superconductivity, which furthermore provide a platform to study fractional vortices.

\section{Acknowledgements}
We thank R.~Arouca, T.~L{\"o}thman, M.~Fogelstr{\"o}m, A.~B.~Vorontsov and E.~Babaev for valuable discussions. We acknowledge M.~Fogelstr{\"o}m, T.~L{\"o}fwander, M.~H\r{a}kansson, O.~Shevtsov, and P.~Stadler for their work on SuperConga. We acknowledge financial support from the European Research Council (ERC) under the European Union Horizon 2020 research and innovation programme (ERC-2017-StG-757553) and the Knut and Alice Wallenberg Foundation through the Wallenberg Academy Fellows program. The computations were enabled by the supercomputing resource Berzelius provided by National Supercomputer Centre (NSC) at Link{\"o}ping University and the Knut and Alice Wallenberg foundation. Additional computations were enabled by resources provided by the National Academic Infrastructure for Supercomputing in Sweden (NAISS) and the Swedish National Infrastructure for Computing (SNIC) at C3SE, HPC2N, and NSC, partially funded by the Swedish Research Council through grant agreements No.~2022-06725 and No.~2018-05973.

\appendix

\section{Coreless vortices in larger systems and opposite chirality}
\label{app:larger_systems}

\begin{figure}[b!]
	\includegraphics[width=\columnwidth]{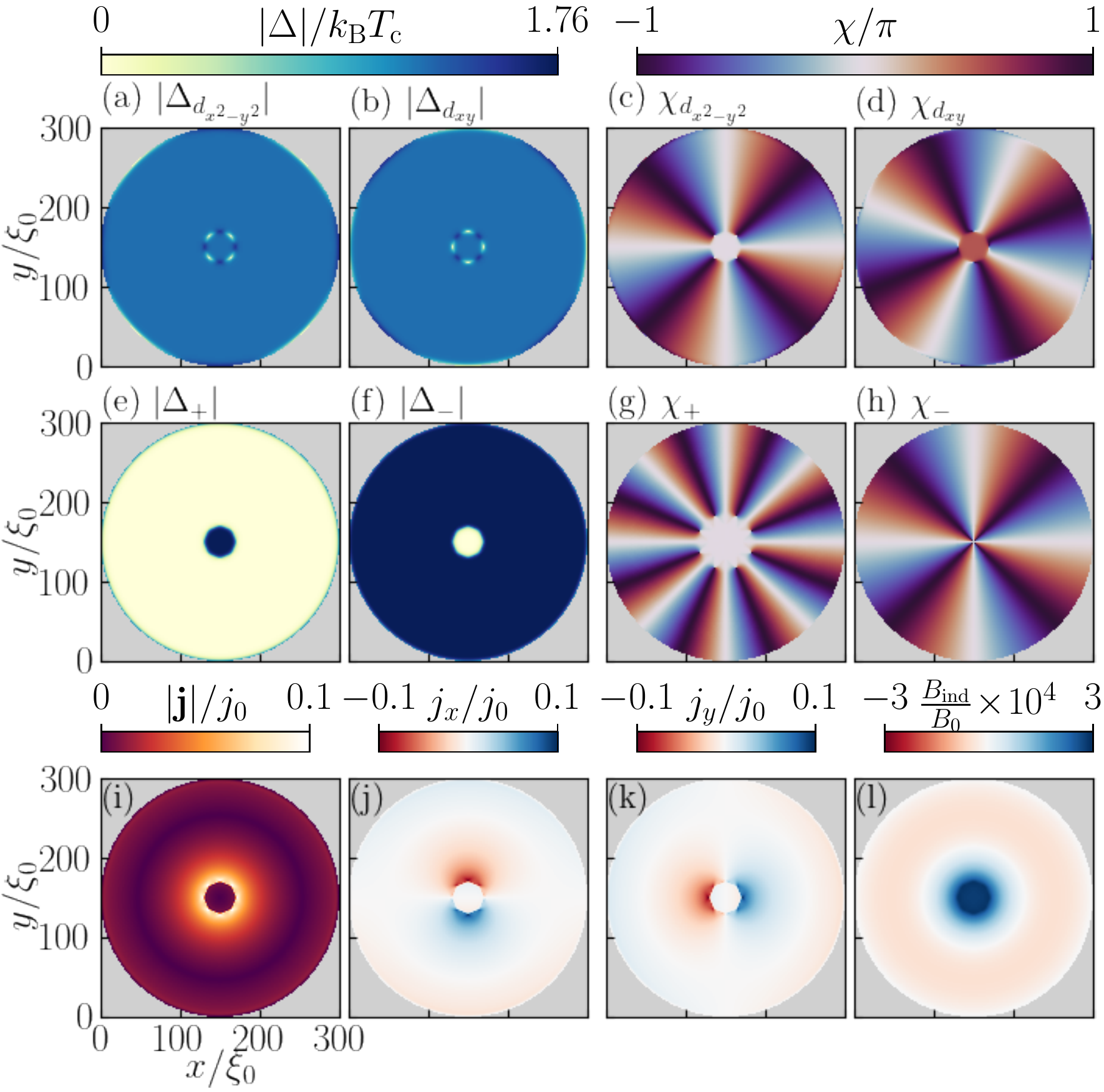}
	\caption{Parallel CV in a disc-shaped system with radius $\mathcal{R}=150\xi_0$, dominant chirality $\Dminus$, $T=0.1\Tc$, $\Phiext=8\Phi_0$, $\lambda=80\xi_0$, $b_0 \equiv 10^{-4}B_0$. First row: amplitudes and phases of the nodal components, second row: same but for chiral components, third row: charge-current density magnitude and $x,y$-components, as well as induced magnetic-flux density. To be compared with Fig.~\ref{fig:parallel_CV}.}
	\label{fig:cv_large_disc}
\end{figure}

In this appendix we show that the qualitative CV features studied in Sec.~\ref{sec:coreless_vortices} remain for much larger systems where the influence of the boundary becomes negligible, and in systems with opposite dominant bulk chirality $\Dminus$. In particular, Fig.~\ref{fig:cv_large_disc} shows various quantities for a parallel CV in a disc with radius $\mathcal{R}=150\xi_0$, dominant bulk chirality $\Dminus$, external flux $\Phiext=8\Phi_0$, temperature $T=0.1\Tc$, and penetration depth $\lambda_0=80\xi_0$, to be compared to Fig.~\ref{fig:parallel_CV}. There is still four well-separated fractional vortices in each nodal component, an octagonal-shaped domain wall in the chiral amplitudes, and integer phase windings $m=-4$ in the dominant phase $\chi_-$ and $p=-8$ in the subdominant phase $\chi_+$ (reversed signs due to reversed bulk chirality). Furthermore, the currents still show the same overall spatial profile and number of sign changes. Interestingly, the additional $p=-8$ phase windings in the subdominant phase $\chi_+$ remains at a small distance outside the CV. We note that for an antiparallel CV in such a large system, the $\pi$-shift in the dominant chirality instead remains closer to the edge (not shown here).

\section{Additional results: interaction with Abrikosov vortices}
\label{app:cv_v_interaction}

\begin{figure*}[bt!]
	\includegraphics[width=\textwidth]{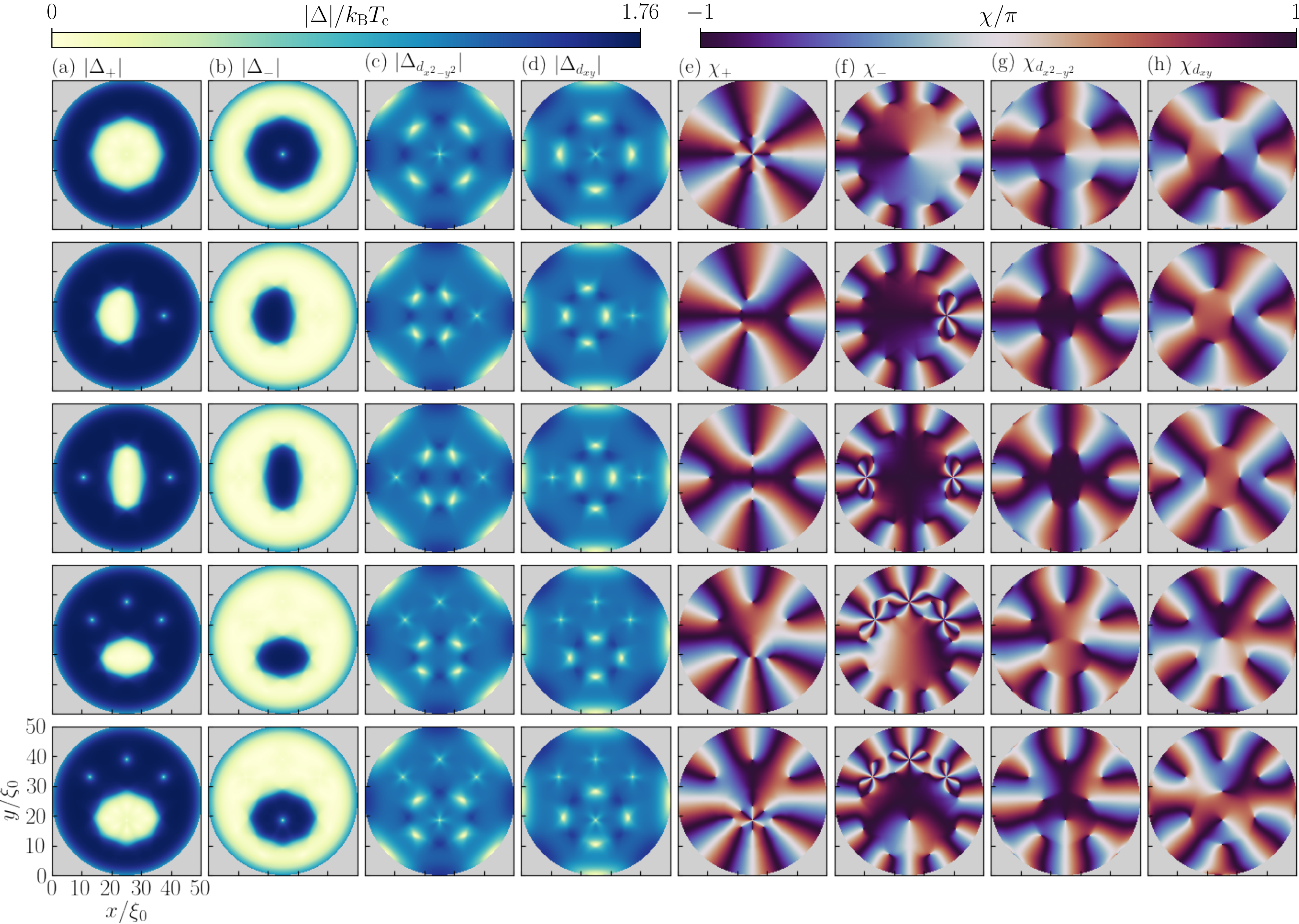}
	\caption{Same as in Fig.~\ref{fig:cv_v:quantities}, but for a parallel CV and without the antivortex scenario.
	}
	\label{fig:cv_v:ASB:quantities}
\end{figure*}

This appendix contains additional numeric results for the interaction between CVs and Abrikosov vortices. In particular, Fig.~\ref{fig:cv_v:ASB:quantities} shows the order parameter amplitudes and phase windings for the same system as in Fig.~\ref{fig:cv_v:ASB:LDOS}, i.e.~this is the analogue of Fig.~\ref{fig:cv_v:quantities} but for a parallel (symmetry-broken) CV caused by negative external flux. Importantly, we note that despite all the additional Abrikosov vortices generating phase windings that overlap with the $p=8$ winding centers of the CV, the latter still generates the distinct shape of the parallel CV studied in the rest of the work.

\bibliographystyle{apsrev4-2}
\bibliography{cite.bib}

\end{document}